\documentclass{aastex63}
\usepackage[utf8]{inputenc}
\usepackage{amsmath}
\usepackage{acronym}

\newcommand{\changed}[1]{{#1}}

\newcommand{\PSR}{PSR\,J1555$-$2908}
\newcommand{\fglsrc}{4FGL\,J1555.7$-$2908}
\newcommand{\atlas}{ATLAS}
\newcommand{\Fermi}{Fermi} 

\graphicspath{{./}{figures/}}

\shorttitle{Discovery of \PSR}
\shortauthors{Ray et al.}

\begin{document}

\newacro{LAT}[LAT]{Large Area Telescope}
\newacro{3FGL}[3FGL]{\Fermi~\acs{LAT} Third Source Catalog}
\newacro{MSP}[MSP]{millisecond pulsar}
\newacro{BIC}[BIC]{Bayesian Information Criterion}

\title{Discovery, Timing, and Multiwavelength Observations of the Black Widow Millisecond Pulsar \PSR}

\correspondingauthor{Paul S. Ray}
\email{paul.ray@nrl.navy.mil}

\author[0000-0002-5297-5278]{Paul S. Ray}
\affiliation{Space Science Division, U.S. Naval Research Laboratory, Washington, DC 20375, USA}
\author[0000-0002-5775-8977]{Lars Nieder}
\affiliation{Albert-Einstein-Institut, Max-Planck-Institut f\"ur Gravitationsphysik, 30167 Hannover, Germany}
\affiliation{Leibniz Universit\"at Hannover, 30167 Hannover, Germany}
\author[0000-0003-4355-3572]{Colin J. Clark}
\affiliation{Jodrell Bank Centre for Astrophysics, Department of Physics and Astronomy, The University of Manchester, M13 9PL, UK}
\affiliation{Albert-Einstein-Institut, Max-Planck-Institut f\"ur Gravitationsphysik, 30167 Hannover, Germany}
\affiliation{Leibniz Universit\"at Hannover, 30167 Hannover, Germany}
\author[0000-0001-5799-9714]{Scott M. Ransom}
\affiliation{National Radio Astronomy Observatory, 520 Edgemont Rd., Charlottesville, VA 22903, USA}
\author{H. Thankful Cromartie}
\affiliation{Cornell Center for Astrophysics and Planetary Science, and Department of Astronomy, Cornell University, Ithaca, NY 14853, USA}
\affiliation{Hubble Fellowship Program Einstein Postdoctoral Fellow, USA}
\author{Dale A. Frail}
\affiliation{National Radio Astronomy Observatory, 1003 Lopezville Road, Socorro, NM 87801, USA}
\author[0000-0002-2557-5180]{Kunal P. Mooley}
\affiliation{National Radio Astronomy Observatory, 1003 Lopezville Road, Socorro, NM 87801, USA}
\affiliation{Caltech, 1200 E. California Blvd. MC 249-17, Pasadena, CA 91125, USA}
\author{Huib Intema} 
\affiliation{International Centre for Radio Astronomy Research (ICRAR), Curtin University, Bentley, WA 6102, Australia}
\affiliation{Leiden Observatory, Leiden University, Niels Bohrweg 2, 2333 CA, Leiden, The Netherlands}
\author{Preshanth Jagannathan}
\affiliation{National Radio Astronomy Observatory, 1003 Lopezville Road, Socorro, NM 87801, USA}
\author{Paul Demorest}
\affiliation{National Radio Astronomy Observatory, 1003 Lopezville Road, Socorro, NM 87801, USA}
\author{Kevin Stovall}
\affiliation{National Radio Astronomy Observatory, 1003 Lopezville Road, Socorro, NM 87801, USA}
\author[0000-0003-4814-2377]{Jules P. Halpern}
\affiliation{Department of Astronomy, Columbia University, 550 West 120th Street, New York, NY 10027, USA}
\author{Julia Deneva}
\altaffiliation{Resident at U.S. Naval Research Laboratory, Washington, DC 20375, USA}
\affiliation{George Mason University, 4400 University Dr, Fairfax, VA 22030, USA}
\author[0000-0002-6449-106X]{Sebastien Guillot}
\affil{IRAP, CNRS, 9 avenue du Colonel Roche, BP 44346, F-31028 Toulouse Cedex 4, France}
\affil{Universit\'{e} de Toulouse, CNES, UPS-OMP, F-31028 Toulouse, France}

\author[0000-0002-0893-4073]{Matthew Kerr}
\affiliation{Space Science Division, U.S. Naval Research Laboratory, Washington, DC 20375, USA}
\author[0000-0003-1699-8867]{Samuel J. Swihart}
\altaffiliation{Resident at U.S. Naval Research Laboratory, Washington, DC 20375, USA}\affiliation{National Research Council Research Associate, National Academy of Sciences, Washington, DC 20001, USA}

\author{Philippe Bruel}
\affiliation{Laboratoire Leprince-Ringuet, \'Ecole polytechnique, CNRS/IN2P3, F-91128 Palaiseau, France}
\author{Ben W.~Stappers}
\affiliation{Jodrell Bank Centre for Astrophysics, Department of Physics and Astronomy, The University of Manchester, M13 9PL, UK}
\author{Andrew Lyne}
\affiliation{Jodrell Bank Centre for Astrophysics, Department of Physics and Astronomy, The University of Manchester, M13 9PL, UK}
\author{Mitch Mickaliger}
\affiliation{Jodrell Bank Centre for Astrophysics, Department of Physics and Astronomy, The University of Manchester, M13 9PL, UK}
\author[0000-0002-1873-3718]{Fernando Camilo}
\affiliation{South African Radio Astronomy Observatory, 2 Fir Street, Observatory 7925, South Africa}
\author[0000-0001-7828-7708]{Elizabeth C. Ferrara}
\affil{Department of Astronomy, University of Maryland, College Park, MD, 20742, USA}
\affil{Center for Exploration and Space Studies (CRESST), NASA/GSFC, Greenbelt, MD 20771, USA}
\affil{NASA Goddard Space Flight Center, Greenbelt, MD 20771, USA}
\author[0000-0002-4013-5650]{Michael T. Wolff}
\affiliation{Space Science Division, U.S. Naval Research Laboratory, Washington, DC 20375, USA}

\author{P.~F.~Michelson}
\affiliation{W. W. Hansen Experimental Physics Laboratory, Kavli Institute for Particle Astrophysics and Cosmology, Department of Physics and SLAC National Accelerator Laboratory, Stanford University, Stanford, CA 94305, USA}

\begin{abstract}
We report the discovery of \PSR, a 1.79~ms radio and gamma-ray pulsar in a 5.6~hr binary system with a minimum companion mass of 0.052~$M_\sun$. This fast and energetic ($\dot E = 3\times10^{35}$ erg~s$^{-1}$) millisecond pulsar was first detected as a gamma-ray point source in \Fermi{} \ac{LAT} sky survey observations. Guided by a steep spectrum radio point source in the \Fermi{} error region, we performed a search at 820 MHz with the Green Bank Telescope that first discovered the pulsations. The initial radio pulse timing observations provided enough information to seed a search for gamma-ray pulsations in the \ac{LAT} data, from which we derive a timing solution valid for the full \Fermi{} mission. In addition to the radio and gamma-ray pulsation discovery and timing, we searched for X-ray pulsations using NICER but no significant pulsations were detected. We also obtained time-series $r$-band photometry that indicates strong heating of the companion star by the pulsar wind.  Material blown off the heated companion eclipses the 820~MHz radio pulse during inferior conjunction of the companion for $\approx10\%$ of the orbit, which is twice the angle subtended by its Roche lobe in an edge-on system.
\end{abstract}


\section{Introduction} 
\label{sec:intro}

One of the great successes of \Fermi{} \ac{LAT} mission \citep{LAT_instrument} 
has been the revelation that a large number of point sources in the GeV gamma-ray sky are powered by pulsars, including large populations both of energetic young pulsars and older, faster spinning millisecond pulsars.
In the early days of the \Fermi{} mission, these discoveries came via two techniques: (1) folding the observed gamma-rays at the pulse period of known radio pulsars \citep[e.g.][]{2009Sci...325..848A} and (2) direct searches for periodicities in the
gamma-ray photon arrival times themselves \citep[e.g.][]{2009Sci...325..840A}. Each of these techniques
had its own biases and selection effects that left many gamma-ray pulsars undiscovered. The next highly successful technique was to use the locations
of \ac{LAT} gamma-ray sources that had pulsar-like characteristics (e.g. curved spectra and low variability) as targets for
directed radio pulsar searches, performed by the \Fermi{} Pulsar Search Consortium (PSC) \citep{PSC}. This led to a bounty of millisecond pulsar discoveries that has continued as the \ac{LAT} survey observations reveal ever-more gamma-ray sources \citep{2011ApJ...727L..16R,2011ApJ...732...47C,2012ApJ...748L...2K,2013ApJ...763L..13R,2013ApJ...773L..12B,2015ApJ...810...85C,2016ApJ...819...34C}.  Most recently, a new technique was employed to reveal pulsar candidates among 
the \ac{LAT} sources: \citet{2016MNRAS.461.1062F} and \citet{2018MNRAS.475..942F} exploited the nearly full-sky 
150 MHz  TIFR GMRT Sky Survey (TGSS) radio survey performed by the Giant Metrewave Radio Telescope (GMRT) to identify steep spectrum radio point sources within the error regions of \ac{LAT} sources. 
These candidates were quickly followed up by the PSC with deep radio searches, leading to several new \ac{MSP} discoveries (initially reported in \citealt{2018MNRAS.475..942F}) and demonstrating the power of this technique. Here we provide details on the discovery of \PSR{} and follow up observations that determined its orbit, discovered gamma-ray pulsations, searched for X-ray pulsations, and made initial studies of its optical companion star.

\section{Image-based Candidate Selection Method}
\label{sec:image}

\PSR\, was initially identified from an image-based search for potential pulsar candidates. 
The method is described in more detail in \citet{2018MNRAS.475..942F} but here we give a 
short summary. We searched for compact, steep spectrum radio sources within the error 
ellipses of unassociated Fermi sources from a preliminary version of the 
\ac{LAT} 8-year catalog \citep[hereafter 4FGL;][]{4FGL}.
We used the GMRT 150 MHz All-Sky Radio Survey \citep[TGSS ADR1;][]{ijmf17} and the 1.4 GHz 
NRAO VLA Sky Survey \citep[NVSS;][]{1998AJ....115.1693C} to calculate initial two-point
spectral indices and source angular diameters for \textit{all} radio sources within \Fermi{}
unassociated sources. Outside of the NVSS survey declination limit ($-45^\circ$), 
but north of the declination limit of TGSS ($-53^\circ$), we used the 843 MHz 
Sydney University Molonglo Sky Survey \citep[SUMSS;][]{1999AJ....117.1578B} 
for the higher frequency of the two-point spectral index. For those rare compact, 
steep spectrum radio sources identified via this method, follow-up interferometric observations at arcsecond resolution were carried out in order to eliminate false positives from the sample, such as high redshift radio galaxies.

This image-based method differs from past searches of \Fermi{} unassociated sources
\citep[e.g.][]{2012ApJ...748L...2K} in that pulsar candidates are selected without
regard to the properties of the \Fermi{} source, such as its spectral shape, temporal 
variability, the size of the error ellipse, or its sky distribution. 
A single unresolved radio source was identified within the error ellipse of P86Y3595 
(now known as \fglsrc{}) with a spectral index $\alpha=-2.5\pm$0.2 (where $\alpha$ is 
given as $S_\nu\propto\nu^\alpha$). The radio source position was measured to be 
R.A.=$15^{\rm h}55^{\rm m}40^{\rm s}\!.69$, decl.=$-29^{\circ}08^{\prime}29^{\prime\prime}\!.0$ 
(J2000) with an uncertainty of about 2\arcsec.

\section{Radio Pulsation Search and Timing Observations}
\label{sec:obs}

As part of a long term effort to discover radio pulsars associated with \Fermi{} sources,
organized by the \Fermi{} PSC, we made targeted observations of the candidates identified 
in \citet{2018MNRAS.475..942F}. We observed P86Y3595 on 2017 February 1 (MJD 57785) for 
30 minutes at 820 MHz with the Robert C. Byrd Green Bank Telescope (GBT).  We used the GUPPI pulsar backend to record 200 MHz of bandwidth with 2048 channels at 61.44 $\mu$s resolution. 
We performed an acceleration search of these data over a range of trial dispersion measures using PRESTO \citep{2002AJ....124.1788R} and identified a strong candidate pulsation with 
1.79 ms period (559.4 Hz) at DM 75.91 pc cm$^{-3}$.  Following the discovery, a 5 minute 
GBT observation at S-band (on MJD 57806) confirmed the discovery and determined the pulse 
width at that frequency to be a very narrow 3\%. 

We also obtained a 30-minute observation on MJD 57788 using the Karl G. Jansky Very Large Array (VLA) in phased-array pulsar mode.  The data were taken while the array was being reconfigured;  20 antennas were in their D-configuration positions (maximum baseline 1~km), and 7 still in A-configuration (maximum baseline 30~km).  The observation was split into three 10-minute scans in order to allow for re-phasing.  Interferometric phases were determined using the nearby calibrator J1554$-$2704, then individual antenna data streams were coherently summed to create a high time resolution single-pixel data stream at the previously-determined radio source position.  The summed voltage data were processed in real time using {\tt dspsr} \citep{dspsr} and recorded using 2048 0.5-MHz channels (1024~MHz total bandwidth, centered on 1500~MHz) at 32~$\mu$s time resolution. 
It is worth noting that this \changed{VLA} observation is quite broadband and exhibits profile evolution across the band. This profile evolution likely biases the measured DM from that observation. 

\begin{figure}
    \centering
    \includegraphics[width=4.0in]{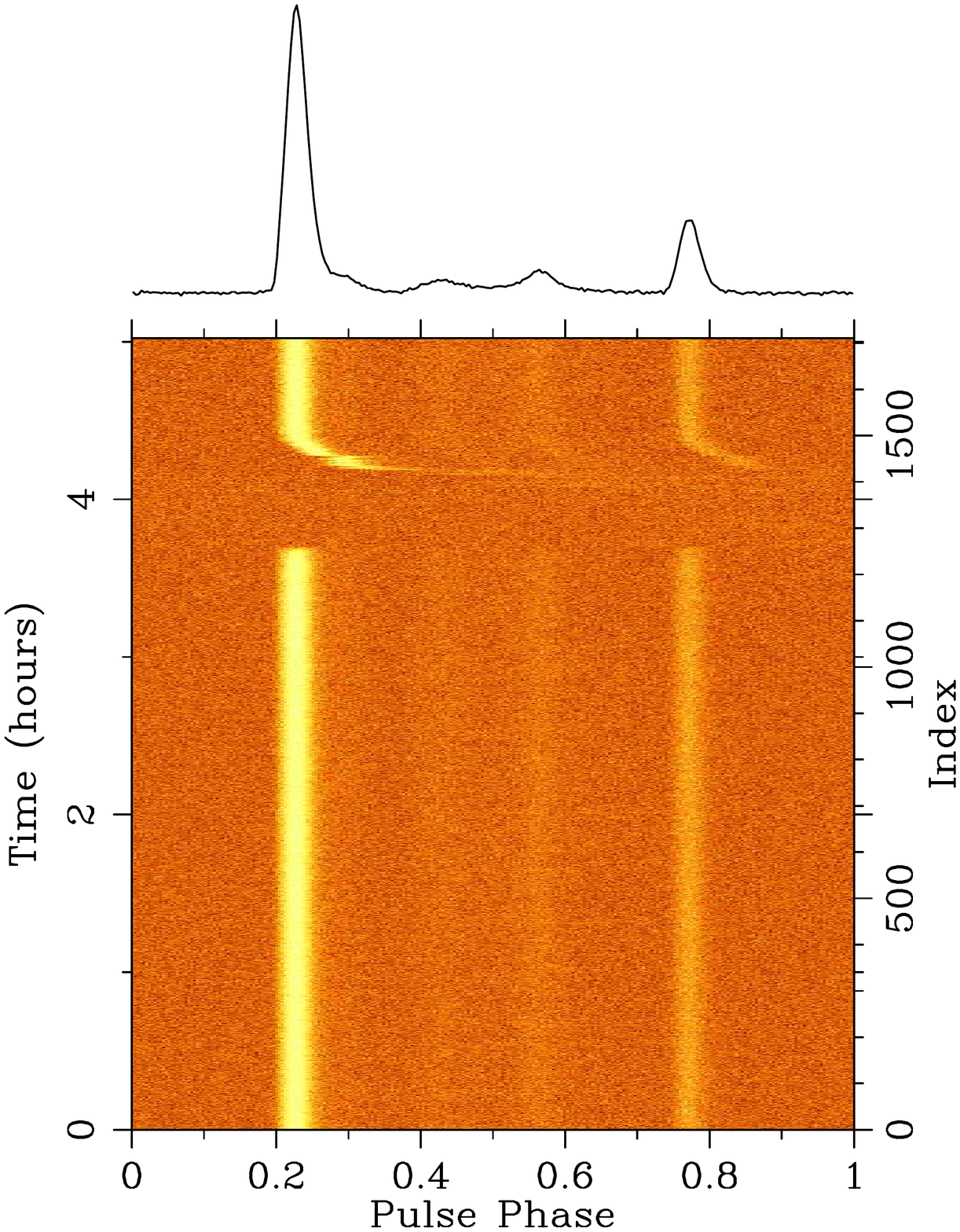}
    \caption{Folded pulse profile (intensity represented by the color) vs. time 
    from 5 hour GBT observation at 820 MHz \added{on MJD 57833}, showing the eclipse ingress and pulse delays during the egress. The summed total intensity pulse profile from the same observation is shown at the top. }
    \label{fig:radio2}
\end{figure}

We also detected the pulsation in a 2-minute GBT observation at 350 MHz that had been taken on MJD 56907 as part of the GBNCC survey \citep{2014ApJ...791...67S}. Lastly we obtained a 5 hour GBT observation on MJD 57833 revealing an eclipse with a sharp ingress and clear pulse delays for about 10 minutes after egress. The eclipse duration is approximately 10\% of the duration of the orbit, and the likely dispersive pulse delays are typical for black-widow pulsar systems.  This observation was taken in coherently-dedispersed search mode, providing a calibrated flux density and high time resolution. A fault in the GUPPI backend at the time of the observation prevented us from extracting reliable polarization measurements from this observation. The eclipse and high S/N total intensity pulse profile from that observation are shown in Figure \ref{fig:radio2}.  

\begin{figure}
    \centering
    \includegraphics[width=5.0in]{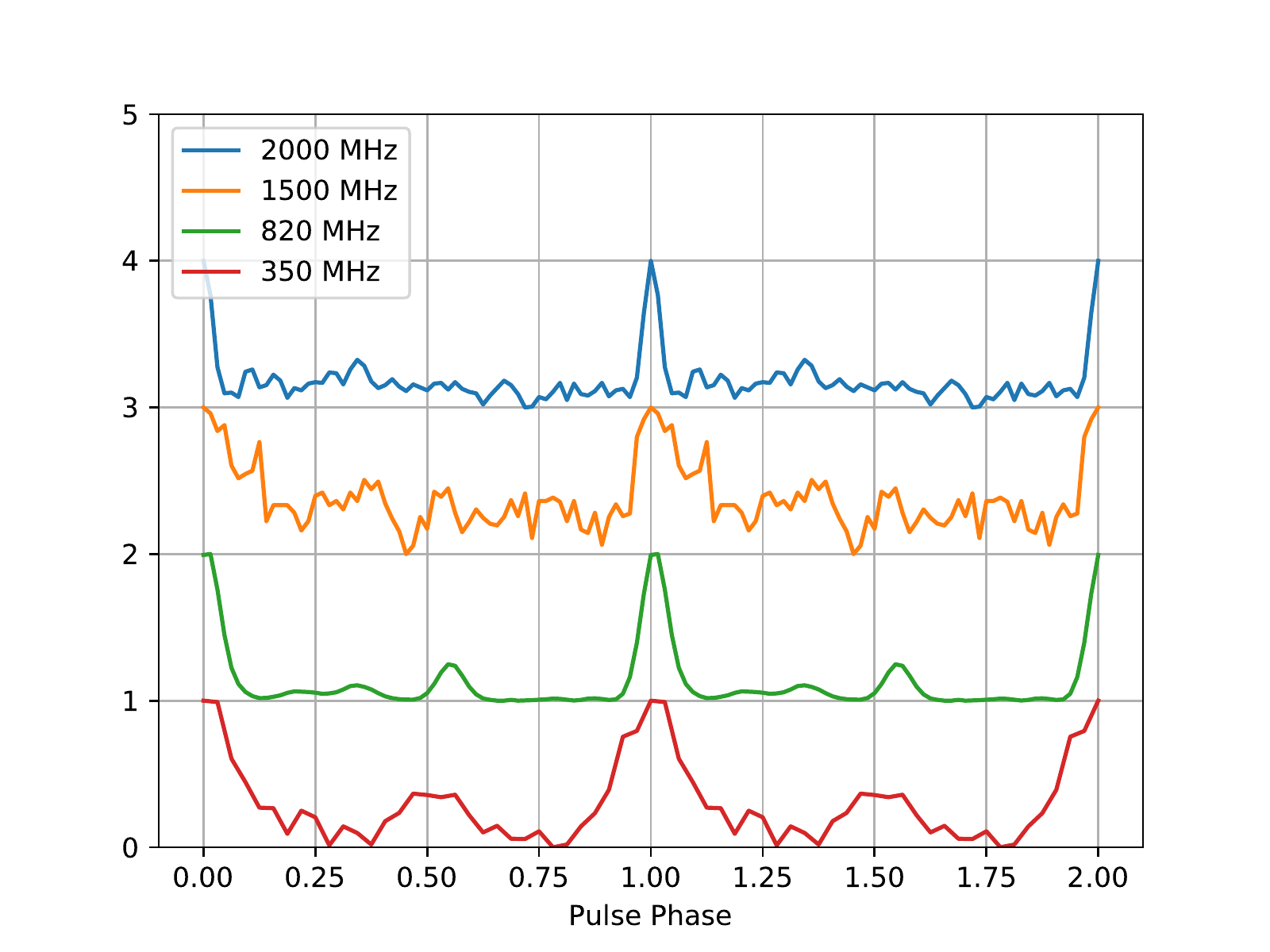}
    \caption{Multi-band radio pulse profiles of \PSR, \changed{from the observations specified in Table \ref{tab:radio} (350 MHz = GBT on MJD 56907, 820 MHz = GBT on 57833, 1500 MHz = VLA on 57788, 2000 MHz = GBT on 57806). The flux density scale is arbitrary and offset vertically for display.} Each profile is aligned to put the peak at phase 0.}
    \label{fig:radio1}
\end{figure}

We also obtained five observations of varying duration using the 76-m Lovell telescope between 2019 November 15 and 2020 September 21. We used a backend based on ROACH FPGA boards \citep{Bassa2016+LEAP} to record coherently de-dispersed, phase-folded data covering a bandwidth of 384 MHz in 0.25 MHz channels centred at 1534 MHz, with 256 phase bins per pulse period. To obtain approximate flux-density calibrations for these observations, we estimated the system equivalent flux density from observations of the Crab nebula on nearby epochs, adjusting for ground spillover due to the low elevation. We estimate the fractional uncertainty on this calibration to be around 20\%. The pulsed flux density was then estimated from each observation by fitting the high S/N 820MHz GBT pulse profile as a template, along with an arbitrary baseline, to the flux-calibrated profiles. The average flux density over the five observations was $S_{1534} = 0.20 \pm 0.05$~mJy.  A log of all the radio observations is presented in Table \ref{tab:radio}. The pulse profile at a range of frequencies is shown in Figure \ref{fig:radio1}.

Combining these radio data, we obtained an initial orbital solution with a period of 5.60 hours and a semimajor axis of 0.151 lt-s.  Because we have only a small amount of radio data, most of the system parameters are better determined by the gamma-ray timing in Section~\ref{s:gamma_pulses}. However, the precise radio TOAs over the 5 hour observation allow us to determine some of the orbital parameters more precisely than is possible from the gamma-ray data alone. In particular, we measure an eccentricity $e = 4.5 \pm 1.5 \times 10^{-6}$, $x = a_1 \sin i = 0.1514468(1)$ lt-s, and an epoch of ascending node of $t_\text{asc} = 57785.53936388(3)$ MJD(TDB). 


\begin{deluxetable}{lrrrrllll}
\tablecaption{Radio Observations of \PSR\label{tab:radio}}
\tablehead{
\colhead{Telescope} & \colhead{MJD} & \colhead{Freq.} & \colhead{BW} & \colhead{Duration} & \colhead{Orbit Phase} &  \colhead{DM} & \colhead{$S_{\nu}$} & \colhead{Notes}\\
                    & \colhead{(UTC)}  & \colhead{(MHz)} & \colhead{(MHz)} & \colhead{(s)} & \colhead{($\phi$)} & \colhead{(pc cm$^{-3}$)} & \colhead{(mJy)}}

\startdata
GBT   & 56907.968322 &  350 &  100 &   120 & 0.668--0.674 & 75.916(2) & & GBNCC \\
GBT   & 57785.510301 &  820 &  200 &  1800 & 0.871--0.959 & 75.9196(6) & &Discovery \\
VLA  & 57788.543634 & 1500 & 1000 &   1920 & 0.862--0.975 & 75.997(4) & &\\ 
GBT   & 57806.554895 & 2000 &  700 &   480 & 0.006--0.030 & 75.91(2) & 0.10(5) &\\
GBT   & 57833.301146 &  820 &  200 & 18000 & 0.562--0.454 & 75.9212(1) & 2.5(5) & Eclipse \\
Lovell & 58802.535248 & 1534 & 384 & 1404 & 0.414--0.483 & 75.94(3) & 0.24(5)\\
Lovell & 58803.504318 & 1534 & 384 & 525  & 0.564--0.590 & 75.95(4) & 0.27(5)\\
Lovell & 58805.433996 & 1534 & 384 & 2397 & 0.828--0.947 & 75.93(2) & 0.16(3)\\
Lovell & 58907.194884 & 1534 & 384 & 907  & 0.661--0.706 & 75.87(4) & 0.19(4)\\
Lovell & 59113.624898 & 1534 & 384 & 1562 & 0.715--0.793 & 75.89(2) & 0.15(3)\\
\enddata
\end{deluxetable}

\section{Gamma-ray Observations}

The initial (not phase-connected) radio timing parameters provided the seed necessary to make a computationally-tractable pulsation search in the gamma-ray data, as we describe here.

\subsection{\Fermi-\ac{LAT} Data Preparation}
To search for gamma-ray pulsations from \PSR{}, we selected Pass 8 \texttt{SOURCE}-class gamma-ray photons \citep{Pass8} detected by the \Fermi{} \ac{LAT} between 2008 August 4 and
2018 April 19 from within a $5^{\circ}$ radius region of interest (RoI)
around the radio position, with energies greater than $100\,$MeV, and with a
maximum zenith angle of $90^{\circ}$.

To increase the sensitivity of the pulsation search, and avoid the need for hard cuts on photon energies and incidence angles, we weighted the contributions of each photon to the pulsation detection statistic \citep{kerr2011}. The weights represent the probability of each photon having been emitted by the targeted gamma-ray source, rather than by a nearby point source, or by the diffuse Galactic or isotropic gamma-ray components. The weights were computed with \texttt{gtsrcprob}, using the \texttt{P8R2\_SOURCE\_V6}  instrument response
functions (IRFs)\footnote{See \url{https://fermi.gsfc.nasa.gov/ssc/data/analysis/LAT_essentials.html}} and a preliminary (``FL8Y'') version of the \Fermi-\ac{LAT} Fourth Source Catalog, and corresponding diffuse and isotropic emission templates, as the input model for the gamma-ray flux within the RoI. 

\subsection{Gamma-ray Pulsation Search and Detection}
\label{s:gamma_pulses}
We performed a gamma-ray pulsation search in $10$ years of \ac{LAT} data around the parameters of a preliminary timing solution that was based on radio data spanning 50 days. A search was necessary as the radio timing parameters were not measured precisely enough to safely extrapolate over multiple years of \ac{LAT} data, and the pulsar's gamma-ray photon flux is too weak to show significant pulsations over such a short time span. For several parameters (spin frequency, orbital period) the precision scales directly with the length of the data span, and at this stage the spin frequency derivative had not been measured.
	
Apart from the radio timing solution, the search parameter space was constrained by a Gaia sky position and the distribution of spin frequency derivatives for known \acp{MSP}. It was assumed that a Gaia source (ID 6041127310076589056) at position R.A.=$15^{\rm h}55^{\rm m}40^{\rm s}\!.65855(7)$, decl.=$-29^{\circ}08^{\prime}28^{\prime\prime}\!.4232(6)$ (J2000) with magnitude $G=20.41$
(EDR3; \citealt{bro21}), coincident with the image-based radio position (Section \ref{sec:image}), is the pulsar's companion star.  In the search, the positional parameters were kept fixed to these values.  Note that this source is too faint for Gaia to measure its proper motion or parallax.  The spin frequency derivative was searched in the range $\dot{f} \in [-2\times10^{-14},0]\,\text{Hz}\,\text{s}^{-1}$, as more than $95\%$ of the known \acp{MSP} fall into this range \citep[ATNF Pulsar Catalogue\footnote{\label{n:atnf}\href{http://www.atnf.csiro.au/research/pulsar/psrcat}{http://www.atnf.csiro.au/research/pulsar/psrcat}};][version 1.64]{manchester2005}. In the case of a non-detection this range would have been extended in steps.
	
The sensitive $H$ statistic \citep{dejager1989,kerr2011} was utilized to search for gamma-ray pulsations. This statistic incoherently combines the Fourier power of the lowest $M$ harmonics. Typically most power is found in the lowest $5$ harmonics \citep{pletsch2014}, so for computational efficiency we truncated the harmonic summing at $M = 5$ as in the successful pulsation search of PSR\,J0952$-$0607 \citep{nieder2019}.
	
An efficient and dense grid covering the parameter space is key to the detection of gamma-ray pulsations. To build such a grid we exploited the distance ``metric'', which is a second-order approximation of the expected fractional loss in squared signal-to-noise ratio due to offsets in the signal parameters \citep{balasubramanian1996,owen1996}. The metric components for the binary pulsar parameters are presented in \citet{Nieder2020+Methods}. Denser grids are required for higher harmonics and thus the grid is built for the highest harmonic, here $M = 5$.
	
The search space was split into smaller parts and carried out in parallel on the \atlas{} computing cluster in Hannover. On one single computer the search would have taken $\sim 70$ days. Distributing the work over $7170$ CPU cores the search only took $\sim 15$ minutes.
	
Significant gamma-ray pulsations were detected over most of the 10-year \ac{LAT} data span used in this search. The resulting pulse phase showed residual time dependence, indicating time evolution of the pulsar spin-down rate. Still, the maximum $H$ statistic detected in the search was $H_{5} = 276.3$. Conservatively assuming that all $6\times10^{11}$ trials were independent \changed{(the actual effective number of trials is smaller)}, the false-alarm probability is $P_{\text{FA}} = 9.6\times10^{-37}$ which confirms the detection of gamma-ray pulsations.

\subsection{Gamma-ray Pulsation Timing}
\label{s:timing}
Following the detection of gamma-ray pulsations, we extended the observation span to cover the most recent \Fermi-\ac{LAT} data \citep{Bruel2018+P305}, up to 2020 August 5 (12 years of data), and using the most recent \texttt{P8R3\_SOURCE\_V3} IRFs. In the 8-yr 4FGL (and the 10-yr ``4FGL-DR2'' iteration, \citealt{4FGL-DR2}), the gamma-ray spectrum of \fglsrc{} is modelled with a simple power law, rather than the curved sub-exponentially-cutoff power-law spectrum typical for gamma-ray pulsars. This is because the curved spectrum did not provide a significantly better fit for the observed gamma-ray flux. This is likely due to a combination of a low overall photon flux, and high uncertainties in the low-energy flux due to the contribution from the diffuse Galactic interstellar emission. \citet{2019A&A...622A.108B} developed a method to obtain optimised photon weights by adjusting the spectral parameters to maximise the resulting weighted $H$ statistic. Adopting this technique to optimise the photon probability weights, we found a pulsar-like sub-exponentially-cutoff power-law spectrum that resulted in a much more significant pulsation detection than was obtained using the simple power-law spectral model. We therefore adopted these weights for the follow-up timing analyses presented below.

\changed{In the timing analysis the pulsar is analyzed precisely using the likelihood, $\mathcal{L}$, and the \aclu{BIC} \citep[\acs{BIC};][]{schwarz1978}.  To measure the pulsar parameters $\boldsymbol{\lambda}$, $\mathcal{L}$ is maximized by fitting a pulse profile to a template pulse profile $\hat{g}$ \citep[see, e.g.,][]{ray2011,kerr2015b,clark2017} by marginalizing over the pulsar parameters and the template parameters jointly as described by \cite{nieder2019}.  The likelihood $\mathcal{L}(\hat{g},\boldsymbol{\lambda}) = \prod_{j=1}^{N} [w_j \hat{g}(\Phi(t_j,\boldsymbol{\lambda})) + (1 - w_j)]$ tests how likely the pulse profile is described via the template $\hat{g}$ versus a flat, noise distribution.  The $j$-th photon contributes according to its weight $w_j$ and the pulsar's phase at emission time $\Phi_j$.  The latter is computed using a phase model $\Phi(t_j,\boldsymbol{\lambda})$, the photon's arrival time at the \ac{LAT} $t_j$, and the pulsar parameters $\boldsymbol{\lambda}$.  An analytic template is constructed as a sum of wrapped Gaussian peaks. For the width we used a log-uniform prior and constrained the range to allow only peaks broader than $1\%$ of a rotation and narrower than half a rotation. All other parameters used a uniform prior.}
	
To account for small phase variations over the full data span, additional spin-frequency derivatives are needed. While only the first derivative was included in the search, four additional derivatives were favored by the \ac{BIC} throughout the timing analysis. Higher orders were disfavored by the \ac{BIC}.

We tested for the presence of additional effects, including proper motion, and eccentricity in the binary system. These parameters were found to be consistent with zero and disfavored by the \ac{BIC}. For those parameters the timing analysis sets $95\%$ confidence upper limits. A circular orbit is favored over an eccentric one with an upper limit on the eccentricity $e^{95\%} < 1.4 \times 10^{-4}$, which is consistent to the value found in the radio timing ($e = 4.5\pm1.5\times10^{-6}$; see Section~\ref{sec:obs}).  The timing analysis clearly favors zero total proper motion $\mu_{\text{t}} = \sqrt{\mu_\alpha^2 \cos^2\delta + \mu_\delta^2}$ setting the upper limit to $\mu_{\text{t}}^{95\%} < 6.4\,\text{mas}\,\text{yr}^{-1}$. The $95\%$ confidence interval on a variable orbital period is $-6.5 \times 10^{-13}\,\text{s}\,\text{s}^{-1} < \dot{P}_{\text{orb}} < 6.1 \times 10^{-13}\,\text{s}\,\text{s}^{-1}$.
	
\begin{figure}
	\centerline{
		\includegraphics[width=0.49\columnwidth]{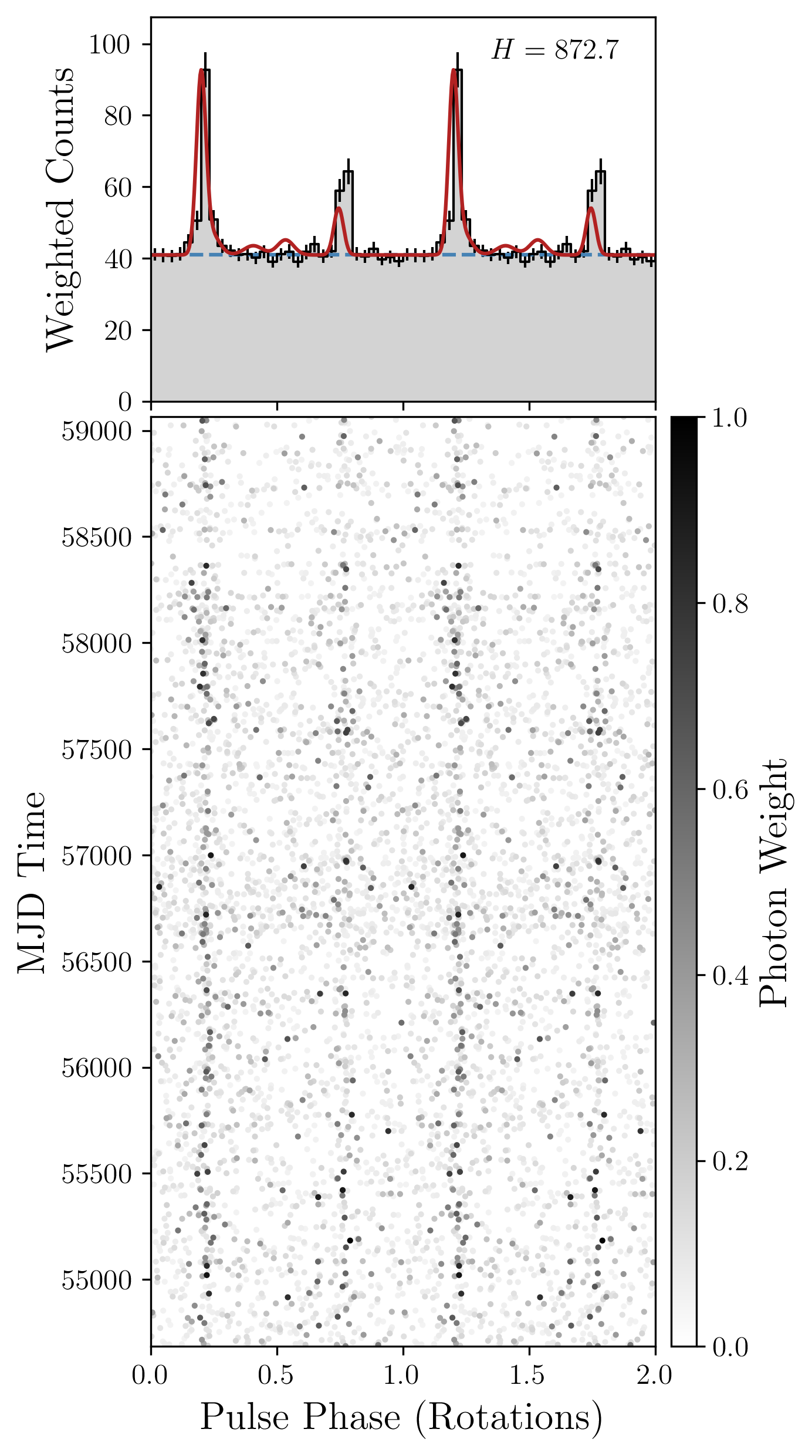}
	}
	\caption{\label{f:timing}
	    Integrated pulse profile and phase-time diagram of \PSR{}, showing two identical rotations for clarity.
		Top: The histogram shows the weighted photon counts with $50$ bins per rotation. The dashed blue line shows the estimated background level $(\sum_j w_j - \sum_j w_j^2) / N_{\rm bins}$, with weights $w_j$ and number of bins $N_{\rm bins}$. The aligned $820$\,MHz radio profile is shown in red.
		Bottom: Each point represents the rotational phase of a detected gamma-ray photon and its gray scale indicates the probability weight.
	}
\end{figure}

Our full timing solution for the 12-year data set is shown in Table~\ref{t:timing} and the gamma-ray phase-time diagram, the gamma-ray pulse profile, and the superposed radio pulse profile are shown in Figure~\ref{f:timing}.  For this final timing analysis, the projected semi-major axis $x$, the epoch of ascending node $t_{\rm{asc}}$, and the two Laplace-Lagrangian parameters $\epsilon_1 = e\sin\omega$ and $\epsilon_2 = e\cos\omega$, \changed{where $\omega$ is the longitude of periastron}, were kept fixed to the values from the radio timing analysis (see Section~\ref{sec:obs}).
	
\begin{deluxetable}{lc}
	\tablecaption{\label{t:timing} Properties of \PSR{} from Gamma-Ray Timing.}
	\tablecolumns{2}
	\tablehead{
        \colhead{Parameter} &
	    \colhead{Value}
    }
	\startdata
	Span of timing data (MJD) \dotfill & $54681$ -- $59066$ \\[0.15em]
	Reference epoch (MJD) \dotfill & $57800.0$ \\[-0.55em]
	\cutinhead{Timing Parameters}
	R.A. (J2000.0)\dotfill & $15^{\rm h}55^{\rm m}40\fs6587(2)$ \\[0.15em]
	Decl. (J2000.0)\dotfill & $-29\arcdeg08\arcmin28\farcs421(8)$ \\[0.15em]
	Spin frequency, $f$ (Hz)\dotfill & $559.44000642609(5)$ \\[0.15em]
	1st spin-frequency derivative, $\dot{f}_\text{obs}$ (Hz s$^{-1}$)\dotfill & $-1.3937(2)\times 10^{-14}$ \\[0.15em]
	2nd spin-frequency derivative, $f^{(2)}_\text{obs}$ (Hz s$^{-2}$)\dotfill & $4(5)\times 10^{-26}$ \\[0.15em]
	3rd spin-frequency derivative, $f^{(3)}_\text{obs}$ (Hz s$^{-3}$)\dotfill & $2(2)\times 10^{-33}$ \\[0.15em]
	4th spin-frequency derivative, $f^{(4)}_\text{obs}$ (Hz s$^{-4}$)\dotfill & $\mathbf{-1.5(7)\times 10^{-40}}$ \\[0.15em]
	5th spin-frequency derivative, $f^{(5)}_\text{obs}$ (Hz s$^{-5}$)\dotfill & $\mathbf{-2.9(8)\times 10^{-48}}$ \\[0.15em]
	Orbital period, $P_\text{orb}$ (day) \dotfill & $0.2335002685(1)$ \\[0.15em]
	Projected semi-major axis$^{\rm a}$, $x$ (lt-s) \dotfill & $0.1514468(1)$ \\[0.15em]
	Epoch of ascending node$^{\rm a}$, $t_\text{asc}$ (MJD) \dotfill & $57785.53936388(3)$ \\[0.15em]
	1st Laplace-Lagrangian parameter$^{\rm a}$, $\epsilon_{1}$ \dotfill & $2(2)\times 10^{-6}$ \\[0.15em]
	2nd Laplace-Lagrangian parameter$^{\rm a}$, $\epsilon_{2}$ \dotfill & $-4(1)\times 10^{-6}$ \\[-0.55em]
	\cutinhead{Derived Properties$^{\rm b}$}
	Spin period, $P_\text{obs}$ (ms) \dotfill & $1.788$ \\[0.15em]
	1st spin-period derivative, $\dot{P}$ (s\,s$^{-1}$) \dotfill & $4.45 \times 10^{-20}$ \\[0.15em]
	Galactic longitude ($l$) \dotfill & $344.\!^\circ48$\\
	Galactic latitude ($b$) \dotfill & $18.\!^\circ50$\\
	Characteristic age, $\tau_\text{c}$ (Gyr) \dotfill & $0.64$ \\[0.15em]
	Spin-down power, $\dot{E}$ (erg s$^{-1}$) \dotfill & $3.1 \times 10^{35}$ \\[0.15em]
	Surface dipole magnetic field, $B_\text{s}$ (G) \dotfill & $2.9 \times 10^{8}$ \\[0.15em]
	Light-cylinder magnetic field, $B_\text{LC}$ (G) \dotfill & $4.6 \times 10^{5}$ \\[0.15em]
	\enddata
	\tablecomments{
	Numbers in parentheses are statistical $1\sigma$ uncertainties on the final digits.  The JPL DE405 solar system ephemeris has been used and times refer to TDB \changed{(using TT = TAI$ + 32.184$ s)}. The timing solution was obtained using the 12-yr \textit{Fermi}-LAT data set described in Section~\ref{s:timing}\\
	$^{\rm a}$ Parameter fixed to radio timing solution, see Section~\ref{sec:obs}.\\
	$^{\rm b}$ Not corrected for Shklovskii and Galactic acceleration effects due to highly uncertain distance measurement.  However, the non-detection of proper motion suggests that these estimates should be accurate to a few percent.
	}
\end{deluxetable}
	
\section{NICER X-ray Pulsation Search}

Although the distance inferred from the dispersion measure is large (see Section~\ref{sec:discussion}), PSR J1555$-$2908 is one of the 10 fastest \acp{MSP} in the Galactic field (i.e., outside of a globular cluster), and the \ac{LAT} timing indicates that the spin down power is very high, at $3.1\times 10^{35}$ erg s$^{-1}$.  This makes it a good candidate to search for X-ray pulsations, as other high $\dot{E}$ \acp{MSP} (e.g. B1937+21, B1821$-$24, and J0218+4232) exhibit bright non-thermal pulsations.  

Motivated by this, we made a 122 ks observation of this source with the NICER X-ray telescope (NICER proposal \#2527), with data accumulated from 2019 May 26 (ObsID 2527010101) through 2019 September 9 (ObsID 2527010142).  \citet{rgr+19} provide a description of the NICER X-ray Timing Instrument and details of the data reduction for millisecond pulsar pulsation searches. We follow a similar procedure here, but since these data were taken with a large angle between the target and the Sun, optical loading was not a significant problem, so we did not mask any of the 52 active detectors. We analyzed our data with HEASoft version 6.27.2 (NICER tools version \texttt{2020-04-23\_V007a}) and updated the gain calibration to version \texttt{nixtiflightpi20170601v006}. Our initial data extraction included energies 0.25--10.0 keV, and made standard data cuts to exclude the SAA and ensure NICER was tracking the source and the source was $>20^\circ$ above the Earth limb. We made no cuts on magnetic cutoff rigidity, or the total count rate, in our initial extraction. This initial extraction yielded 117.8 ks of good time.  For each photon, we computed the pulse phase using \texttt{photonphase} from PINT \citep{PINT} and the timing model provided in Table \ref{t:timing}.


To exclude high background regions, instead of making an arbitrary cut on count rate, we developed a tool \texttt{ni\_Htest\_sortgti.py} that divides the data into segments of no more than 100 s (and no less than 10 s). These segments are sorted by mean count rate. Since the count rate from the pulsar is very low and presumably constant, this is equivalent to sorting by background rate. Then the $H$-test detection statistic is evaluated cumulatively, going from the lowest background to the highest. This algorithm thus finds the optimal background rate cut that maximizes the S/N of any detected signal. The script repeats this process over a large grid of $E_\mathrm{min}$ and $E_\mathrm{max}$ energy cuts to search for an optimal energy band as well. For our detection statistic, we chose to use the $H$-test \citep{dejager1989}, which is preferred to the $Z^2$ test for unknown pulse profiles that may be very sharp, as is seen in the non-thermal X-ray MSPs. 

For \PSR{}, this procedure did not reveal any very strong candidate pulsation. The highest $H$-test found corresponded to a single-trial significance of 3.05~$\sigma$, when searching the range 0.26--2.74 keV, with the algorithm selecting the lowest background 106.7 ks out of the total. In those data, the mean count rate was 0.80~s$^{-1}$. Since there were a large number of trials over the energy and background cuts, this does not represent a significant detection.

In the 0.25--2.0 keV band, the median count rate (source+background) was 0.75~s$^{-1}$, with 90\% of segments below 1.13~s$^{-1}$.  In the 2--8 keV band, the median count rate (source+background) was 0.23~s$^{-1}$, with 90\% of segments $<0.6$~s$^{-1}$.

\section{Optical Counterpart}

\begin{figure}
  \centerline{
  \includegraphics[width=\columnwidth]{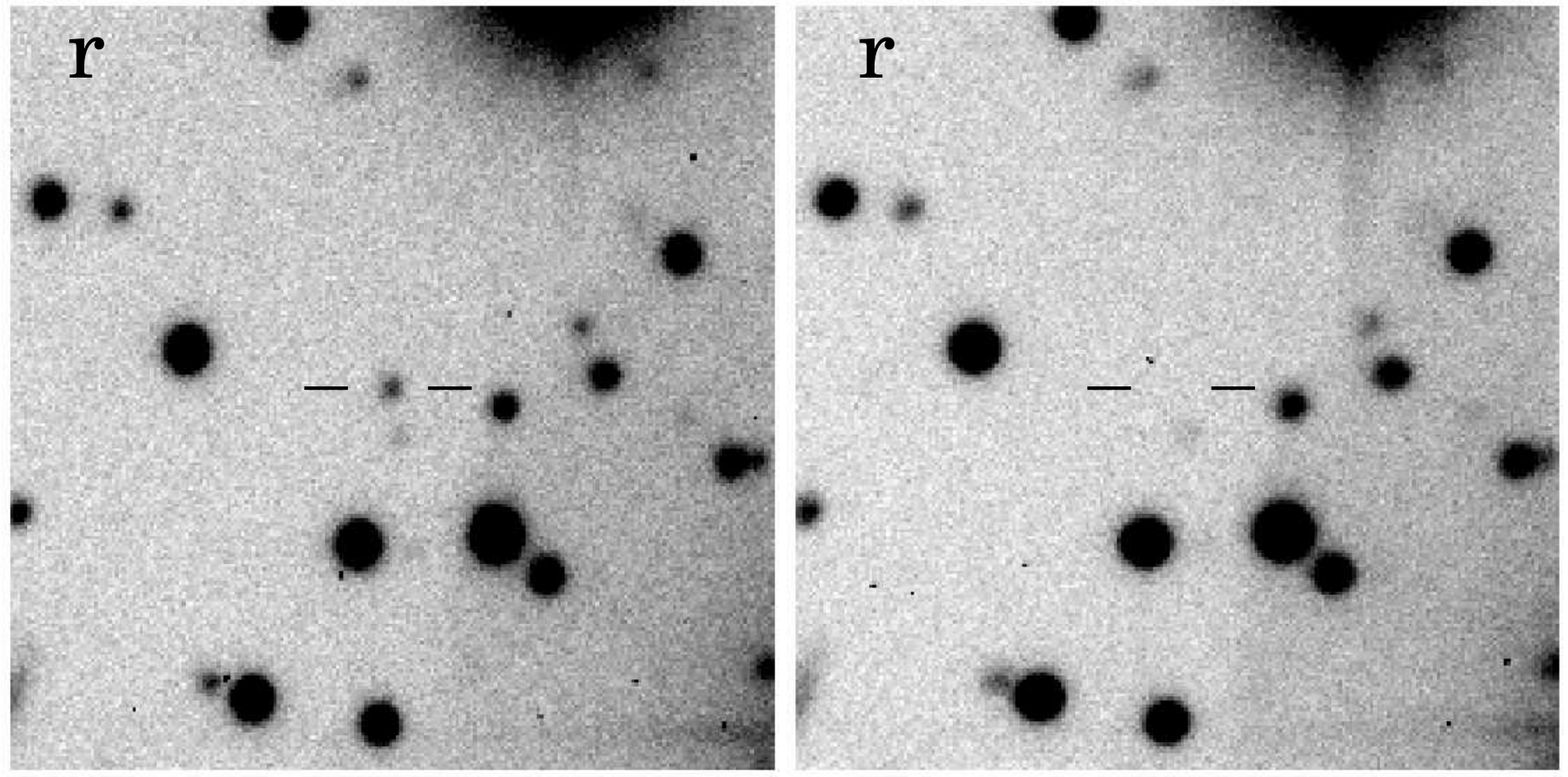}
}
\caption{MDM 2.4 m $r$-band images centered on \PSR, near orbital brightness maximum (left) and minimum (undetected, right).  The field displayed is $60^{\prime\prime}\times60^{\prime\prime}$; north is up and east is to the left.  
}
\label{f:finding_chart}
\end{figure}

We performed differential time-series photometry at the position of \PSR\ using the 2.4-m Hiltner telescope of the MDM Observatory on 2018 June 12 and 15.  A back-illuminated SITe CCD with $1024\times1024$ $24\,\mu$ pixels, each subtending $0.\!^{\prime\prime}275$, was exposed through an SDSS $r$ filter.  Integrations of 5 minutes each were obtained for a total of 7.16 hours over the two nights.  The differential photometry was calibrated using a comparison star from Pan-STARRS.  A variable star with a maximum brightness of $r=20.4$ and displaying a typical heating-dominated light curve was detected at the image-based radio position (Section \ref{sec:image}), also consistent with the Gaia counterpart position and magnitude (Section \ref{s:gamma_pulses}) and the gamma-ray timing position (Table \ref{t:timing}).
A finding chart is shown in Figure~\ref{f:finding_chart}.

\begin{figure}

\centerline{
  \includegraphics[width=1.1\columnwidth]{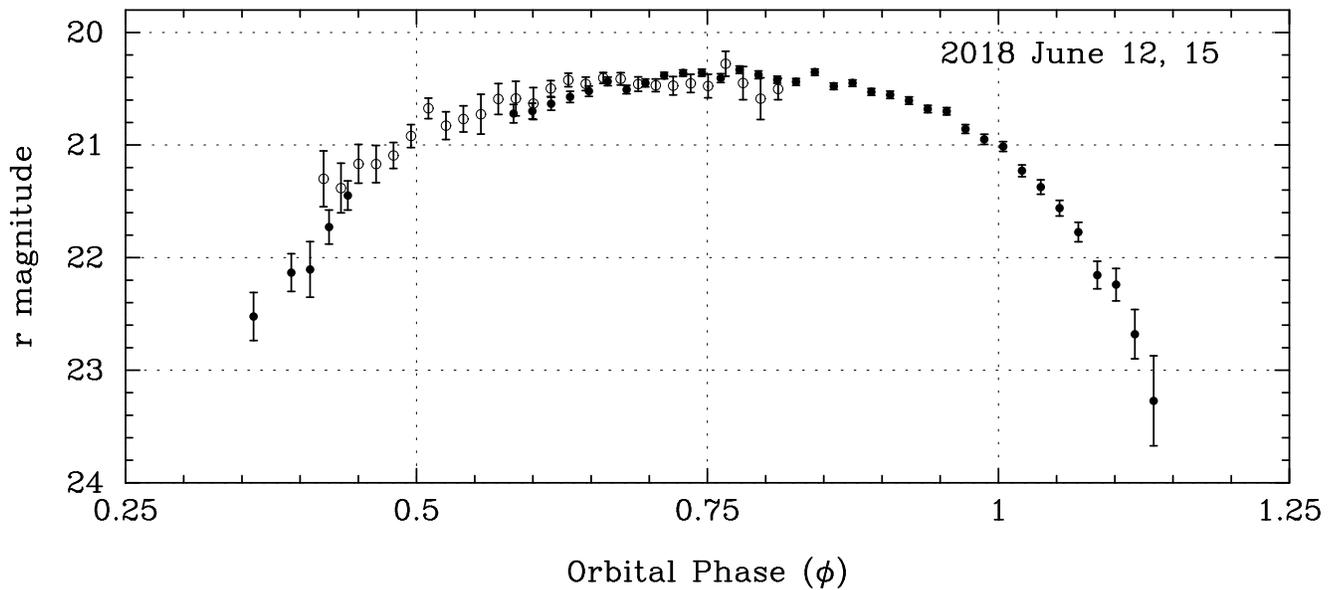}
}
\vspace{-6.6in}
\caption{MDM 2.4 m $r$-band light curve of \PSR\ phased according to the ephemeris of Table~\ref{t:timing}. Filled circles are from 2018 June 12 and open circles are from 2018 June 15.  Differential photometry was calibrated using a comparison star from Pan-STARRS.  Upper limits of $r>23$ close to inferior conjunction of the companion star ($\phi=0.25\pm0.1$) are not shown.  The data quality is worse on June 15 because of poor seeing.}
\label{f:light_curve}
\end{figure}

It was immediately apparent that the timescale and pattern of variability are consistent with orbital modulation of a low-mass companion star heated by the pulsar wind.  The phased light curve using the orbital ephemeris of Table~\ref{t:timing} is shown in Figure~\ref{f:light_curve}.  As expected, the peak brightness occurs near phase 0.75, superior conjunction of the companion, where we are viewing the heated side of the star. The ``night'' side of the companion was not detected; only upper limits of $r>23$ were obtained at phases $\phi=0.25\pm0.1$.

\section{Discussion}
\label{sec:discussion}

The radio discovery of \PSR{} continues  the trend of successfully finding radio pulsations from MSPs by targeting LAT sources. It also demonstrates
the power of using imaging radio surveys to discover steep-spectrum point sources
associated with \ac{LAT} sources.  These are prime pulsar candidates.

The pulsar discovery provides some constraints on the distance to the source. The dispersion measure of 75.9 pc~cm$^{-3}$ gives distance estimates of 2.65 kpc for the NE2001 electron density model \citep{NE2001} and 7.55 kpc for the YMW16 model \citep{YMW16}.
The difference of the two estimates can be traced to the properties of the thick disk component in the electron density models.  The DM at high Galactic latitude is dominated by the thick disk, which in the YMW16 model has a smaller central density and a larger scale height than in the NE2001 model, and an integrated column density that is slightly more than half that of NE2001.  High-latitude pulsars are thus assigned a larger DM distance in YMW16.  \PSR\ is at Galactic coordinates $(\ell,b)=(344.\!^{\circ}5,+18.\!^{\circ}5)$, and has the largest DM of neighboring pulsars within a $5^{\circ}$ radius.  This pushes it to a large, albeit uncertain distance in YMW16.

From our flux-calibrated observations with the Lovell telescope, we estimated an L-band pulsed flux density of $S_{1534} = 0.20 \pm 0.05$ mJy. Comparing this to the imaging flux density measured at 150 MHz by \citet{2018MNRAS.475..942F} of $S_{150} = 133.9 \pm 14.4$ mJy, and assuming a simple power-law spectrum $S_{\nu} \propto \nu^{\alpha}$, yields a two-point spectral index estimate of $\alpha = -2.8 \pm 0.1$, confirming the steep spectrum. 

\PSR{} is a fast and energetic pulsar, currently the 13th fastest galactic MSP in the ATNF Pulsar Catalogue. The small upper limit on the measured proper motion of 6.4 mas yr$^{-1}$ sets an upper limit on the Shklovskii correction to the spindown rate of at most 3\%. The larger of the two DM distance estimates gives an intrinsic spindown luminosity of $2.98 \times 10^{35}$ erg s$^{-1}$. This can be compared with the measured gamma-ray luminosity. The associated LAT source in the 4FGL DR3 catalog is 4FGL J1555.7$-$2908 with an energy flux above 100 MeV (G100) of $(4.66 \pm 0.61) \times 10^{-12}$ erg cm$^{-2}$ s$^{-1}$. This corresponds to an efficiency of converting spindown energy into $>100$ MeV gamma-rays of 1.3\% for $d=2.65$ kpc and 11\% for $d=7.55$ kpc.
These are both plausible values for gamma-ray MSPs \citep[see][]{2PC}. Alternatively, \citet{2PC} define a rough heuristic relationship $L_\gamma^h = \sqrt{10^{33} \dot{E}}$~erg~s$^{-1}$, which gives an expected gamma-ray luminosity of $1.8 \times 10^{34}$ erg~s$^{-1}$. This implies a distance of 5.7 kpc, but with substantial uncertainty based on the observed scatter around that relationship. 

The X-ray pulsation upper limits can be compared to the population of gamma-ray millisecond pulsars whose properties are compiled in \citet{2PC}.  The ratio of $>100$ MeV gamma-ray flux ($G_{100}$) to non-thermal X-ray flux ($F_X^\mathrm{nt}$; unabsorbed 0.3--10 keV) ranges over at least 2 orders of magnitude, from $\sim 20$ to over several thousand. If \PSR{} is like the energetic MSP PSR B1937+21 (essentially the most optimistic case), which has a $G_{100}/F_X^{nt} = 23$, then $F_X$ could be as high as $3 \times 10^{-13}$ erg s$^{-1}$. Assuming $n_{\rm H} = 2.2\times 10^{21}$~cm$^{-2}$ scaled from the DM per \citet{HeNgKaspi2013} and a power-law index of 2.0, yields a NICER count rate of 0.044~s$^{-1}$.  The detectability of pulsations depends on the unknown pulse shape (ranging from the worst case of a pure sinusoid to a very narrow, low duty cycle pulse), but we can estimate the detection significance using the method of \citet{1987A&A...175..353B}.  For this flux, we should have detected these pulsations if the source pulsed fraction was larger than 75\% (for a sinusoid) or 40\% (for a narrow pulse). For less favorable gamma-ray to X-ray flux ratios, e.g. with predicted count rates $<0.018$~s$^{-1}$, our observation would not have detected pulsations even with very narrow pulse shapes. Consequently, the X-ray observation is not sensitive enough to place very stringent constraints on the X-ray luminosity, but could have detected pulsations if this pulsar had an X-ray luminosity and pulse shape similar to PSR B1937+21.

In long term timing of millisecond pulsars, for example with pulsar timing arrays like NANOGrav, it is very unusual for frequency derivatives above $\dot{f}$ to be significantly detected. One notable counterexample is PSR J1024$-$0719, which has a significantly detected second spin frequency derivative \citep{Kaplan2016,Bassa2016}, ascribed to a long period orbit. A discussion of the measured timing noise and the possibility of PSR J1555$-$2908 being in a triple system with a long period outer orbit will be presented in Nieder et al. (2021, in prep.).

Another unusual trait of this MSP is the aligned radio and gamma-ray peaks,
with similar morphology. These MSPs, dubbed Type A (for ``aligned'')
by \citet{Espinoza2013} or Class II by \citet{Johnson2014}, account for only a few
percent of the known gamma-ray MSPs in the Fermi Third Pulsar Catalog (3PC, in prep). This subset of MSPs are generally highly energetic, but the strongest correlation seems to be with the inferred magnetic field at the light cylinder ($B_\mathrm{LC}$; \citealt{Espinoza2013}). In the radio band they also generally have steeper spectra and lower levels of linear polarization than the general MSP population, and several are known to emit giant pulses.
PSR J1555$-$2908 follows this trend, with its high $\dot{E}$, very large $B_\mathrm{LC}$, and steep ($\alpha = -2.5$; \citealt{2018MNRAS.475..942F}) radio spectral index; however, we do not see evidence for giant pulses in the long 820 MHz GBT observation. Unfortunately, an instrumental issue with the GUPPI backend prevented us from making a measurement of the linear polarization fraction. The most successful models for these pulsars  invoke co-located radio and gamma-ray emission regions in the outer magnetosphere, such as altitude-limited Two Pole Caustic models (alTPC; \citealt{Johnson2014}).





\begin{figure}
    \centering
    \includegraphics[width=4.0in]{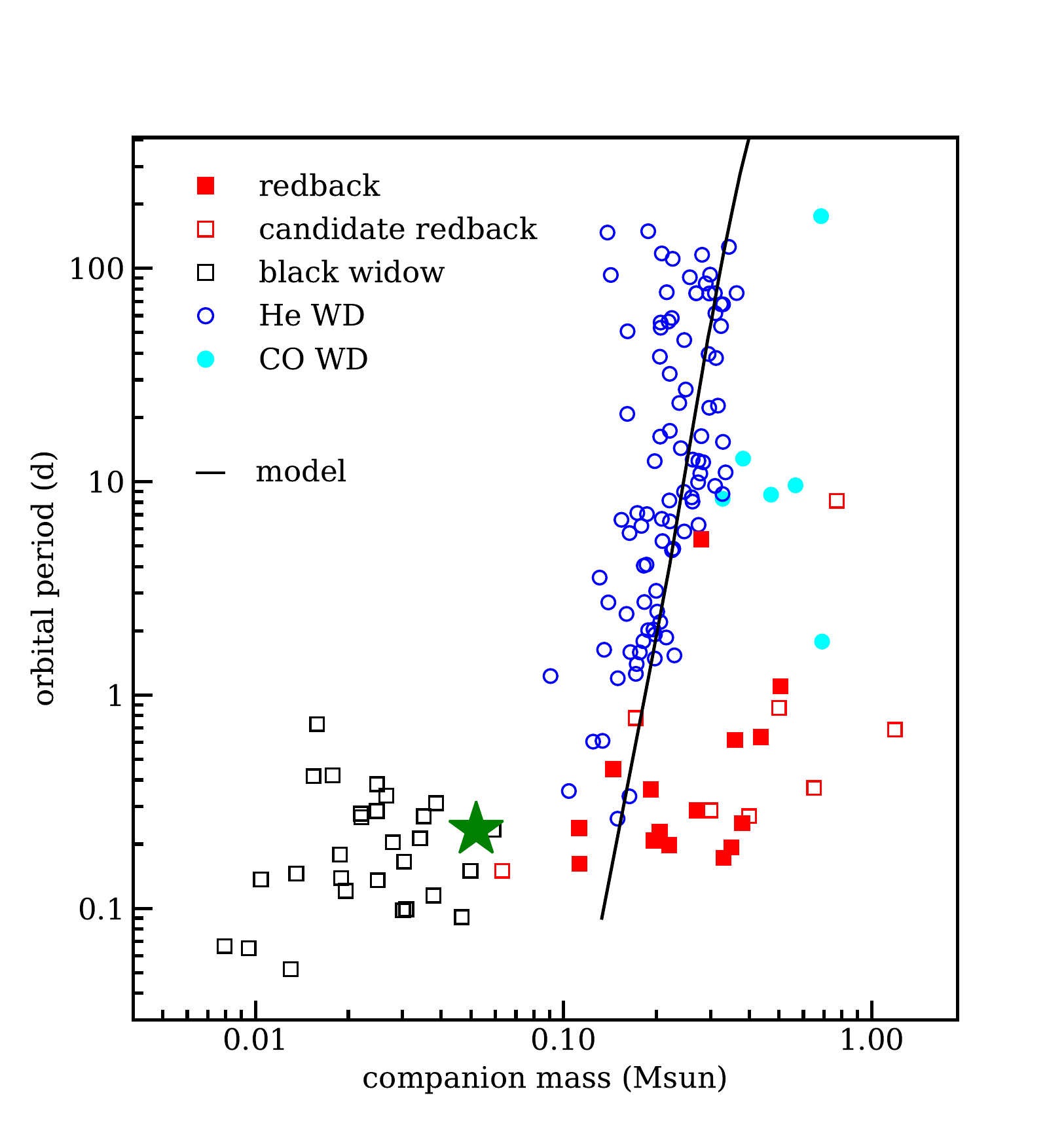}
    \caption{Companion mass vs. orbital period for recycled field MSPs with known companion star types. Most MSPs have He white dwarfs (open blue circles) from binary evolution, well-represented by models from \citet{TS99}. These models (black line) assume an initial secondary mass of 1.0 $M_\sun$, and denote the endpoints of an ensemble of systems with varying initial period, not the evolution of a single binary. The few CO white dwarfs (filled cyan circles) had close common-envelope evolution. The field redbacks (filled/open red squares) and black widows (open black squares) are visible at short orbital period. PSR J1555$-$2908 is marked with a green star at its minimum companion mass assuming an edge-on orbit and a neutron star mass of 1.4 $M_\sun$. Figure and caption adapted from \citet{Strader2019}\label{fig:spiders}.
}
    \label{fig:my_label}
\end{figure}

As has been the case for many of the MSPs discovered in searches of LAT sources, this pulsar is in an interacting binary system. The orbital solution gives a minimum companion mass of $0.052 M_\odot$, for an assumed neutron star mass of 1.4 $M_\odot$. The very low mass companion [much lower than the white dwarf mass predicted by \citet{TS99} of 0.17 $M_\odot$] and radio eclipses put it in the black widow class. As shown in Figure \ref{fig:spiders}, the minimum companion mass is relatively large for MSPs in this orbital period range, which tend to have minimum companion masses of 0.02--0.03 $M_\odot$, but the eclipse around phase 0.25 indicates that the system is nearly edge-on, so the true companion mass is close to the minimum value.  The eclipse spans $\approx10\%$ of the orbit, which is twice the width of the companion's Roche lobe (see below) in an edge-on system, consistent with absorption by an evaporated wind.  The measured eccentricity is very low, but in family for black widow type systems.

Our optical observations confirm that the optical source detected by Gaia is the heated companion star to \PSR{}. The amplitude of the modulation, likely more than 4 magnitudes from maximum to the non-detection at minimum, is among the largest seen in black widow companions \citep[e.g.,][]{Breton2012,Draghis2019+BWs}, which is consistent with heating being powered by the unusually high spin-down power.
With a light curve in only one filter, and incomplete detection around the orbit, it is not possible to constrain a detailed model of the system geometry and irradiation of the companion star by the pulsar wind. Instead, we make a back-of-the-envelope calculation to check whether the peak observed flux of the companion is consistent with heating supplied by the spin-down power of the pulsar, and see if this constrains the temperature of the heated side of the companion.  For an assumed isotropic pulsar wind and a Roche-lobe filling companion, the maximum
irradiating power can be approximated as $L_{\rm irr}=\eta\,\dot E\,r_L^2/4 a^2$, where $r_L$ is the radius of the Roche lobe of the companion, $a$ is the orbital separation, $\eta$ is the heating efficiency, and $\dot E\approx 3.1\times10^{35}$ erg~s$^{-1}$ for \PSR.  Assuming masses of $1.4\,M_{\odot}$ and $0.052\,M_{\odot}$, the ratio $r_L/a=0.154$ according to \citet{egg83}.  The maximum irradiating power is then $L_{\rm irr}=1.8\times10^{33}\eta$ erg~s$^{-1}$.

In order to compare this to the observed optical flux at orbital phase 0.75, we assume an orbital inclination of $i\approx90^{\circ}$, which is justified by the radio eclipses and the large amplitude of optical modulation around the orbit.  We also correct the observed $r=20.4$ for $A_r=0.316$, the maximum extinction along the line of sight \citep{sch11}.  Converting magnitude to flux density using $f^{\rm Vega}_{\lambda,\rm eff}=2.40\times10^{-9}$ erg~cm$^{-2}$~s$^{-1}$~\AA$^{-1}$ \citep{fuk95}, the absorption-corrected flux density at $\lambda^{\rm Vega}_{\rm eff}=6205$~\AA\ is $f_{\lambda,{\rm eff}}=2.2\times10^{-17}$ erg~cm$^{-2}$~s$^{-1}$~\AA$^{-1}$. For the DM distance we assume either 2.65~kpc or 7.55~kpc according to the two models of the electron density distribution referenced above.

At $d=2.65$~kpc, we find that the temperature of the heated half of the star is required to be $\ge 5,000$\,K in order for the star to fit within its Roche lobe.  At $T_{\rm h}=5,000$\,K, the efficiency $\eta$ would be $0.05$.  For higher temperature the heated area would decrease, allowing the star to partially fill its Roche lobe, and the efficiency would increase modestly. Assuming instead that $d=7.55$~kpc, the hot side must have $T_{\rm h}>9,000$~K to fit within the Roche-lobe, but the efficiency becomes $\eta\ge0.44$. Any temperature above 14,000\,K would require $\eta>1$.  Thus, the range of plausible models would be much more limited at the larger distance. 

We have additionally obtained 12~hr of simultaneous photometric observations in the $u$, $g$ and $i$ bands with the high-speed multi-band imager ULTRACAM on the 3.6m New Technology Telescope at ESO La Silla, as well as 0.65 orbits of optical spectroscopy, covering the optical maximum, with X-SHOOTER on ESO's Very Large Telescope. These observations and their modelling will be presented in Kennedy et al. (2021, in prep.).





\acknowledgments

We thank Nicholas C.\ S.\ Ray (West Potomac High School) for his careful 
screening of the candidates from the GBT pulsation search that first 
revealed the pulsar.

The \textit{Fermi} LAT Collaboration acknowledges generous ongoing support
from a number of agencies and institutes that have supported both the
development and the operation of the LAT as well as scientific data analysis.
These include the National Aeronautics and Space Administration and the
Department of Energy in the United States, the Commissariat \`a l'Energie Atomique
and the Centre National de la Recherche Scientifique / Institut National de Physique
Nucl\'eaire et de Physique des Particules in France, the Agenzia Spaziale Italiana
and the Istituto Nazionale di Fisica Nucleare in Italy, the Ministry of Education,
Culture, Sports, Science and Technology (MEXT), High Energy Accelerator Research
Organization (KEK) and Japan Aerospace Exploration Agency (JAXA) in Japan, and
the K.~A.~Wallenberg Foundation, the Swedish Research Council and the
Swedish National Space Board in Sweden.

Additional support for science analysis during the operations phase is gratefully
acknowledged from the Istituto Nazionale di Astrofisica in Italy and the Centre
National d'\'Etudes Spatiales in France. This work performed in part under DOE
Contract DE-AC02-76SF00515.

Fermi work at NRL is supported by NASA. ECF supported by NASA under award number
80GSFC21M0002. Pulsar research at Jodrell Bank Centre for Astrophysics and access 
to the Lovell telescope is supported by a consolidated grant from the UK Science 
and Technology Facilities Council (STFC). 

MDM Observatory is operated by Dartmouth College, Columbia University, Ohio State University, Ohio University, and the University of Michigan.
The National Radio Astronomy Observatory and the Green Bank Observatory are facilities of the National Science Foundation operated under cooperative agreement by Associated Universities, Inc.

S.M.R. is a CIFAR Fellow and is supported by the NSF Physics Frontiers Center awards 1430284 and 2020265.

C.J.C. acknowledges support from the ERC under the European Union's Horizon 2020 research and innovation programme (grant agreement No.715051; Spiders).

This research has made use of data and/or 
software provided by the High Energy Astrophysics Science Archive Research 
Center (HEASARC), which is a service of the Astrophysics Science Division at 
NASA/GSFC and the High Energy Astrophysics Division of the Smithsonian 
Astrophysical Observatory.  This 
research has made use of the NASA Astrophysics Data System (ADS) and the 
arXiv.


\vspace{5mm}
\facilities{Fermi, GBT, VLA, NICER}


\software{\texttt{astropy} \citep{Astropy2}, 
\texttt{PINT} (\citealt{PINT}; \url{https://github.com/nanograv/pint}),
HEAsoft (\texttt{ascl:1408.004}; \url{https://heasarc.gsfc.nasa.gov/docs/software/heasoft/})
}

\bibliographystyle{aasjournal}
\bibliography{main}

\begin{thebibliography}{}
\expandafter\ifx\csname natexlab\endcsname\relax\def\natexlab#1{#1}\fi
\providecommand{\url}[1]{\href{#1}{#1}}
\providecommand{\dodoi}[1]{doi:~\href{http://doi.org/#1}{\nolinkurl{#1}}}
\providecommand{\doeprint}[1]{\href{http://ascl.net/#1}{\nolinkurl{http://ascl.net/#1}}}
\providecommand{\doarXiv}[1]{\href{https://arxiv.org/abs/#1}{\nolinkurl{https://arxiv.org/abs/#1}}}

\bibitem[{{Abdo} {et~al.}(2009{\natexlab{a}}){Abdo}, {Ackermann}, {Ajello},
  {Atwood}, {Axelsson}, {Baldini}, {Ballet}, {Barbiellini}, {Baring},
  {Bastieri}, {Baughman}, {Bechtol}, {Bellazzini}, {Berenji}, {Bignami},
  {Blandford}, {Bloom}, {Bonamente}, {Borgland}, {Bregeon}, {Brez}, {Brigida},
  {Bruel}, {Burnett}, {Caliandro}, {Cameron}, {Camilo}, {Caraveo}, {Carlson},
  {Casand jian}, {Cecchi}, {{\c{C}}elik}, {Charles}, {Chekhtman}, {Cheung},
  {Chiang}, {Ciprini}, {Claus}, {Cognard}, {Cohen-Tanugi}, {Cominsky},
  {Conrad}, {Corbet}, {Cutini}, {Dermer}, {Desvignes}, {de Angelis}, {de Luca},
  {de Palma}, {Digel}, {Dormody}, {do Couto e Silva}, {Drell}, {Dubois},
  {Dumora}, {Edmonds}, {Farnier}, {Favuzzi}, {Fegan}, {Focke}, {Frailis},
  {Freire}, {Fukazawa}, {Funk}, {Fusco}, {Gargano}, {Gasparrini}, {Gehrels},
  {Germani}, {Giebels}, {Giglietto}, {Giordano}, {Glanzman}, {Godfrey},
  {Grenier}, {Grondin}, {Grove}, {Guillemot}, {Guiriec}, {Hanabata}, {Harding},
  {Hayashida}, {Hays}, {Hobbs}, {Hughes}, {J{\'o}hannesson}, {Johnson},
  {Johnson}, {Johnson}, {Johnson}, {Johnston}, {Kamae}, {Katagiri}, {Kataoka},
  {Kawai}, {Kerr}, {Kn{\"o}dlseder}, {Kocian}, {Kramer}, {Kuss}, {Lande},
  {Latronico}, {Lemoine-Goumard}, {Longo}, {Loparco}, {Lott}, {Lovellette},
  {Lubrano}, {Madejski}, {Makeev}, {Manchester}, {Marelli}, {Mazziotta},
  {McConville}, {McEnery}, {McLaughlin}, {Meurer}, {Michelson}, {Mitthumsiri},
  {Mizuno}, {Moiseev}, {Monte}, {Monzani}, {Morselli}, {Moskalenko}, {Murgia},
  {Nolan}, {Norris}, {Nuss}, {Ohsugi}, {Omodei}, {Orlando}, {Ormes}, {Paneque},
  {Panetta}, {Parent}, {Pelassa}, {Pepe}, {Pesce-Rollins}, {Piron}, {Porter},
  {Rain{\`o}}, {Rando}, {Ransom}, {Ray}, {Razzano}, {Rea}, {Reimer}, {Reimer},
  {Reposeur}, {Ritz}, {Rochester}, {Rodriguez}, {Romani}, {Roth}, {Ryde},
  {Sadrozinski}, {Sanchez}, {Sander}, {Saz Parkinson}, {Scargle}, {Schalk},
  {Sgr{\`o}}, {Siskind}, {Smith}, {Smith}, {Spandre}, {Spinelli}, {Stappers},
  {Starck}, {Striani}, {Strickman}, {Suson}, {Tajima}, {Takahashi}, {Tanaka},
  {Thayer}, {Thayer}, {Theureau}, {Thompson}, {Thorsett}, {Tibaldo}, {Torres},
  {Tosti}, {Tramacere}, {Uchiyama}, {Usher}, {Van Etten}, {Vasileiou},
  {Venter}, {Vilchez}, {Vitale}, {Waite}, {Wallace}, {Wang}, {Watters}, {Webb},
  {Weltevrede}, {Winer}, {Wood}, {Ylinen}, \& {Ziegler}}]{2009Sci...325..848A}
{Abdo}, A.~A., {Ackermann}, M., {Ajello}, M., {et~al.} 2009{\natexlab{a}},
  Science, 325, 848, \dodoi{10.1126/science.1176113}

\bibitem[{{Abdo} {et~al.}(2009{\natexlab{b}}){Abdo}, {Ackermann}, {Ajello},
  {Anderson}, {Atwood}, {Axelsson}, {Baldini}, {Ballet}, {Barbiellini},
  {Baring}, {Bastieri}, {Baughman}, {Bechtol}, {Bellazzini}, {Berenji},
  {Bignami}, {Blandford}, {Bloom}, {Bonamente}, {Borgland}, {Bregeon}, {Brez},
  {Brigida}, {Bruel}, {Burnett}, {Caliandro}, {Cameron}, {Caraveo},
  {Casandjian}, {Cecchi}, {{\c{C}}elik}, {Chekhtman}, {Cheung}, {Chiang},
  {Ciprini}, {Claus}, {Cohen-Tanugi}, {Conrad}, {Cutini}, {Dermer}, {de
  Angelis}, {de Luca}, {de Palma}, {Digel}, {Dormody}, {do Couto e Silva},
  {Drell}, {Dubois}, {Dumora}, {Farnier}, {Favuzzi}, {Fegan}, {Fukazawa},
  {Funk}, {Fusco}, {Gargano}, {Gasparrini}, {Gehrels}, {Germani}, {Giebels},
  {Giglietto}, {Giommi}, {Giordano}, {Glanzman}, {Godfrey}, {Grenier},
  {Grondin}, {Grove}, {Guillemot}, {Guiriec}, {Gwon}, {Hanabata}, {Harding},
  {Hayashida}, {Hays}, {Hughes}, {J{\'o}hannesson}, {Johnson}, {Johnson},
  {Johnson}, {Kamae}, {Katagiri}, {Kataoka}, {Kawai}, {Kerr}, {Kn{\"o}dlseder},
  {Kocian}, {Kuss}, {Land e}, {Latronico}, {Lemoine-Goumard}, {Longo},
  {Loparco}, {Lott}, {Lovellette}, {Lubrano}, {Madejski}, {Makeev}, {Marelli},
  {Mazziotta}, {McConville}, {McEnery}, {Meurer}, {Michelson}, {Mitthumsiri},
  {Mizuno}, {Monte}, {Monzani}, {Morselli}, {Moskalenko}, {Murgia}, {Nolan},
  {Norris}, {Nuss}, {Ohsugi}, {Omodei}, {Orlando}, {Ormes}, {Paneque},
  {Parent}, {Pelassa}, {Pepe}, {Pesce-Rollins}, {Pierbattista}, {Piron},
  {Porter}, {Primack}, {Rain{\`o}}, {Rando}, {Ray}, {Razzano}, {Rea}, {Reimer},
  {Reimer}, {Reposeur}, {Ritz}, {Rochester}, {Rodriguez}, {Romani}, {Ryde},
  {Sadrozinski}, {Sanchez}, {Sander}, {Parkinson}, {Scargle}, {Sgr{\`o}},
  {Siskind}, {Smith}, {Smith}, {Spand re}, {Spinelli}, {Starck}, {Strickman},
  {Suson}, {Tajima}, {Takahashi}, {Takahashi}, {Tanaka}, {Thayer}, {Thompson},
  {Tibaldo}, {Tibolla}, {Torres}, {Tosti}, {Tramacere}, {Uchiyama}, {Usher},
  {Van Etten}, {Vasileiou}, {Vilchez}, {Vitale}, {Waite}, {Wang}, {Watters},
  {Winer}, {Wolff}, {Wood}, {Ylinen}, {Ziegler}, \& {Fermi LAT
  Collaboration}}]{2009Sci...325..840A}
---. 2009{\natexlab{b}}, Science, 325, 840, \dodoi{10.1126/science.1175558}

\bibitem[{{Abdo} {et~al.}(2013){Abdo}, {Ajello}, {Allafort}, {Baldini},
  {Ballet}, {Barbiellini}, {Baring}, {Bastieri}, {Belfiore}, {Bellazzini}, \&
  et~al.}]{2PC}
{Abdo}, A.~A., {Ajello}, M., {Allafort}, A., {et~al.} 2013, \apjs, 208, 17,
  \dodoi{10.1088/0067-0049/208/2/17}

\bibitem[{{Abdollahi} {et~al.}(2020){Abdollahi}, {Acero}, {Ackermann},
  {Ajello}, {Atwood}, {Axelsson}, {Baldini}, {Ballet}, {Barbiellini},
  {Bastieri}, {Becerra Gonzalez}, {Bellazzini}, {Berretta}, {Bissaldi}, {Bland
  ford}, {Bloom}, {Bonino}, {Bottacini}, {Brandt}, {Bregeon}, {Bruel},
  {Buehler}, {Burnett}, {Buson}, {Cameron}, {Caputo}, {Caraveo}, {Casandjian},
  {Castro}, {Cavazzuti}, {Charles}, {Chaty}, {Chen}, {Cheung}, {Chiaro},
  {Ciprini}, {Cohen-Tanugi}, {Cominsky}, {Coronado-Bl{\'a}zquez}, {Costantin},
  {Cuoco}, {Cutini}, {D'Ammando}, {DeKlotz}, {de la Torre Luque}, {de Palma},
  {Desai}, {Digel}, {Di Lalla}, {Di Mauro}, {Di Venere}, {Dom{\'\i}nguez},
  {Dumora}, {Fana Dirirsa}, {Fegan}, {Ferrara}, {Franckowiak}, {Fukazawa},
  {Funk}, {Fusco}, {Gargano}, {Gasparrini}, {Giglietto}, {Giommi}, {Giordano},
  {Giroletti}, {Glanzman}, {Green}, {Grenier}, {Griffin}, {Grondin}, {Grove},
  {Guiriec}, {Harding}, {Hayashi}, {Hays}, {Hewitt}, {Horan},
  {J{\'o}hannesson}, {Johnson}, {Kamae}, {Kerr}, {Kocevski}, {Kovac'evic'},
  {Kuss}, {Landriu}, {Larsson}, {Latronico}, {Lemoine-Goumard}, {Li},
  {Liodakis}, {Longo}, {Loparco}, {Lott}, {Lovellette}, {Lubrano}, {Madejski},
  {Maldera}, {Malyshev}, {Manfreda}, {Marchesini}, {Marcotulli},
  {Mart{\'\i}-Devesa}, {Martin}, {Massaro}, {Mazziotta}, {McEnery}, {Mereu},
  {Meyer}, {Michelson}, {Mirabal}, {Mizuno}, {Monzani}, {Morselli},
  {Moskalenko}, {Negro}, {Nuss}, {Ojha}, {Omodei}, {Orienti}, {Orlando},
  {Ormes}, {Palatiello}, {Paliya}, {Paneque}, {Pei}, {Pe{\~n}a-Herazo},
  {Perkins}, {Persic}, {Pesce-Rollins}, {Petrosian}, {Petrov}, {Piron}, {Poon},
  {Porter}, {Principe}, {Rain{\`o}}, {Rando}, {Razzano}, {Razzaque}, {Reimer},
  {Reimer}, {Remy}, {Reposeur}, {Romani}, {Saz Parkinson}, {Schinzel},
  {Serini}, {Sgr{\`o}}, {Siskind}, {Smith}, {Spandre}, {Spinelli}, {Strong},
  {Suson}, {Tajima}, {Takahashi}, {Tak}, {Thayer}, {Thompson}, {Tibaldo},
  {Torres}, {Torresi}, {Valverde}, {Van Klaveren}, {van Zyl}, {Wood},
  {Yassine}, \& {Zaharijas}}]{4FGL}
{Abdollahi}, S., {Acero}, F., {Ackermann}, M., {et~al.} 2020, \apjs, 247, 33,
  \dodoi{10.3847/1538-4365/ab6bcb}

\bibitem[{{Astropy Collaboration} {et~al.}(2018){Astropy Collaboration},
  {Price-Whelan}, {Sip{\H o}cz}, {G{\"u}nther}, {Lim}, {Crawford}, {Conseil},
  {Shupe}, {Craig}, {Dencheva}, {Ginsburg}, {VanderPlas}, {Bradley},
  {P{\'e}rez-Su{\'a}rez}, {de Val-Borro}, {Aldcroft}, {Cruz}, {Robitaille},
  {Tollerud}, {Ardelean}, {Babej}, {Bach}, {Bachetti}, {Bakanov}, {Bamford},
  {Barentsen}, {Barmby}, {Baumbach}, {Berry}, {Biscani}, {Boquien}, {Bostroem},
  {Bouma}, {Brammer}, {Bray}, {Breytenbach}, {Buddelmeijer}, {Burke},
  {Calderone}, {Cano Rodr{\'{\i}}guez}, {Cara}, {Cardoso}, {Cheedella},
  {Copin}, {Corrales}, {Crichton}, {D'Avella}, {Deil}, {Depagne}, {Dietrich},
  {Donath}, {Droettboom}, {Earl}, {Erben}, {Fabbro}, {Ferreira}, {Finethy},
  {Fox}, {Garrison}, {Gibbons}, {Goldstein}, {Gommers}, {Greco}, {Greenfield},
  {Groener}, {Grollier}, {Hagen}, {Hirst}, {Homeier}, {Horton}, {Hosseinzadeh},
  {Hu}, {Hunkeler}, {Ivezi{\'c}}, {Jain}, {Jenness}, {Kanarek}, {Kendrew},
  {Kern}, {Kerzendorf}, {Khvalko}, {King}, {Kirkby}, {Kulkarni}, {Kumar},
  {Lee}, {Lenz}, {Littlefair}, {Ma}, {Macleod}, {Mastropietro}, {McCully},
  {Montagnac}, {Morris}, {Mueller}, {Mumford}, {Muna}, {Murphy}, {Nelson},
  {Nguyen}, {Ninan}, {N{\"o}the}, {Ogaz}, {Oh}, {Parejko}, {Parley}, {Pascual},
  {Patil}, {Patil}, {Plunkett}, {Prochaska}, {Rastogi}, {Reddy Janga},
  {Sabater}, {Sakurikar}, {Seifert}, {Sherbert}, {Sherwood-Taylor}, {Shih},
  {Sick}, {Silbiger}, {Singanamalla}, {Singer}, {Sladen}, {Sooley},
  {Sornarajah}, {Streicher}, {Teuben}, {Thomas}, {Tremblay}, {Turner},
  {Terr{\'o}n}, {van Kerkwijk}, {de la Vega}, {Watkins}, {Weaver}, {Whitmore},
  {Woillez}, {Zabalza}, \& {Astropy Contributors}}]{Astropy2}
{Astropy Collaboration}, {Price-Whelan}, A.~M., {Sip{\H o}cz}, B.~M., {et~al.}
  2018, \aj, 156, 123, \dodoi{10.3847/1538-3881/aabc4f}

\bibitem[{{Atwood} {et~al.}(2012){Atwood}, {Albert}, {Baldini}, {Tinivella},
  {Bregeon}, {Pesce-Rollins}, {Sgr{\`o}}, {Bruel}, {Charles}, {Drlica-Wagner},
  {Franckowiak}, {Jogler}, {Rochester}, {Usher}, {Wood}, {Cohen-Tanugi}, \&
  {S.~Zimmer for the Fermi-LAT Collaboration}}]{Pass8}
{Atwood}, W., {Albert}, A., {Baldini}, L., {et~al.} 2012, in Proceedings of the
  4th Fermi Symposium, ed. T.~J. {Brandt}, N.~{Omodei}, \& C.~{Wilson-Hodge},
  eConf C121028, 8

\bibitem[{{Atwood} {et~al.}(2009){Atwood}, {Abdo}, {Ackermann}, {Althouse},
  {Anderson}, {Axelsson}, {Baldini}, {Ballet}, {Band}, {Barbiellini},
  {Bartelt}, {Bastieri}, {Baughman}, {Bechtol}, {B{\'e}d{\'e}r{\`e}de},
  {Bellardi}, {Bellazzini}, {Berenji}, {Bignami}, {Bisello}, {Bissaldi},
  {Blandford}, {Bloom}, {Bogart}, {Bonamente}, {Bonnell}, {Borgland},
  {Bouvier}, {Bregeon}, {Brez}, {Brigida}, {Bruel}, {Burnett}, {Busetto},
  {Caliandro}, {Cameron}, {Caraveo}, {Carius}, {Carlson}, {Casandjian},
  {Cavazzuti}, {Ceccanti}, {Cecchi}, {Charles}, {Chekhtman}, {Cheung},
  {Chiang}, {Chipaux}, {Cillis}, {Ciprini}, {Claus}, {Cohen-Tanugi},
  {Condamoor}, {Conrad}, {Corbet}, {Corucci}, {Costamante}, {Cutini}, {Davis},
  {Decotigny}, {DeKlotz}, {Dermer}, {de Angelis}, {Digel}, {do Couto e Silva},
  {Drell}, {Dubois}, {Dumora}, {Edmonds}, {Fabiani}, {Farnier}, {Favuzzi},
  {Flath}, {Fleury}, {Focke}, {Funk}, {Fusco}, {Gargano}, {Gasparrini},
  {Gehrels}, {Gentit}, {Germani}, {Giebels}, {Giglietto}, {Giommi}, {Giordano},
  {Glanzman}, {Godfrey}, {Grenier}, {Grondin}, {Grove}, {Guillemot}, {Guiriec},
  {Haller}, {Harding}, {Hart}, {Hays}, {Healey}, {Hirayama}, {Hjalmarsdotter},
  {Horn}, {Hughes}, {J{\'o}hannesson}, {Johansson}, {Johnson}, {Johnson},
  {Johnson}, {Johnson}, {Kamae}, {Katagiri}, {Kataoka}, {Kavelaars}, {Kawai},
  {Kelly}, {Kerr}, {Klamra}, {Kn{\"o}dlseder}, {Kocian}, {Komin}, {Kuehn},
  {Kuss}, {Landriu}, {Latronico}, {Lee}, {Lee}, {Lemoine-Goumard}, {Lionetto},
  {Longo}, {Loparco}, {Lott}, {Lovellette}, {Lubrano}, {Madejski}, {Makeev},
  {Marangelli}, {Massai}, {Mazziotta}, {McEnery}, {Menon}, {Meurer},
  {Michelson}, {Minuti}, {Mirizzi}, {Mitthumsiri}, {Mizuno}, {Moiseev},
  {Monte}, {Monzani}, {Moretti}, {Morselli}, {Moskalenko}, {Murgia},
  {Nakamori}, {Nishino}, {Nolan}, {Norris}, {Nuss}, {Ohno}, {Ohsugi}, {Omodei},
  {Orlando}, {Ormes}, {Paccagnella}, {Paneque}, {Panetta}, {Parent}, {Pearce},
  {Pepe}, {Perazzo}, {Pesce-Rollins}, {Picozza}, {Pieri}, {Pinchera}, {Piron},
  {Porter}, {Poupard}, {Rain{\`o}}, {Rando}, {Rapposelli}, {Razzano}, {Reimer},
  {Reimer}, {Reposeur}, {Reyes}, {Ritz}, {Rochester}, {Rodriguez}, {Romani},
  {Roth}, {Russell}, {Ryde}, {Sabatini}, {Sadrozinski}, {Sanchez}, {Sander},
  {Sapozhnikov}, {Parkinson}, {Scargle}, {Schalk}, {Scolieri}, {Sgr{\`o}},
  {Share}, {Shaw}, {Shimokawabe}, {Shrader}, {Sierpowska-Bartosik}, {Siskind},
  {Smith}, {Smith}, {Spandre}, {Spinelli}, {Starck}, {Stephens}, {Strickman},
  {Strong}, {Suson}, {Tajima}, {Takahashi}, {Takahashi}, {Tanaka}, {Tenze},
  {Tether}, {Thayer}, {Thayer}, {Thompson}, {Tibaldo}, {Tibolla}, {Torres},
  {Tosti}, {Tramacere}, {Turri}, {Usher}, {Vilchez}, {Vitale}, {Wang},
  {Watters}, {Winer}, {Wood}, {Ylinen}, \& {Ziegler}}]{LAT_instrument}
{Atwood}, W.~B., {Abdo}, A.~A., {Ackermann}, M., {et~al.} 2009, \apj, 697,
  1071, \dodoi{10.1088/0004-637X/697/2/1071}

\bibitem[{{Balasubramanian} {et~al.}(1996){Balasubramanian}, {Sathyaprakash},
  \& {Dhurandhar}}]{balasubramanian1996}
{Balasubramanian}, R., {Sathyaprakash}, B.~S., \& {Dhurandhar}, S.~V. 1996,
  \prd, 53, 3033, \dodoi{10.1103/PhysRevD.53.3033}

\bibitem[{{Ballet} {et~al.}(2020){Ballet}, {Burnett}, {Digel}, \&
  {Lott}}]{4FGL-DR2}
{Ballet}, J., {Burnett}, T.~H., {Digel}, S.~W., \& {Lott}, B. 2020, arXiv
  e-prints, arXiv:2005.11208.
\newblock \doarXiv{2005.11208}

\bibitem[{{Bassa} {et~al.}(2016{\natexlab{a}}){Bassa}, {Janssen},
  {Karuppusamy}, {Kramer}, {Lee}, {Liu}, {McKee}, {Perrodin}, {Purver},
  {Sanidas}, {Smits}, \& {Stappers}}]{Bassa2016+LEAP}
{Bassa}, C.~G., {Janssen}, G.~H., {Karuppusamy}, R., {et~al.}
  2016{\natexlab{a}}, \mnras, 456, 2196, \dodoi{10.1093/mnras/stv2755}

\bibitem[{{Bassa} {et~al.}(2016{\natexlab{b}}){Bassa}, {Janssen}, {Stappers},
  {Tauris}, {Wevers}, {Jonker}, {Lentati}, {Verbiest}, {Desvignes}, {Graikou},
  {Guillemot}, {Freire}, {Lazarus}, {Caballero}, {Champion}, {Cognard},
  {Jessner}, {Jordan}, {Karuppusamy}, {Kramer}, {Lazaridis}, {Lee}, {Liu},
  {Lyne}, {McKee}, {Os{\l}owski}, {Perrodin}, {Sanidas}, {Shaifullah}, {Smits},
  {Theureau}, {Tiburzi}, \& {Zhu}}]{Bassa2016}
{Bassa}, C.~G., {Janssen}, G.~H., {Stappers}, B.~W., {et~al.}
  2016{\natexlab{b}}, \mnras, 460, 2207, \dodoi{10.1093/mnras/stw1134}

\bibitem[{{Bhattacharyya} {et~al.}(2013){Bhattacharyya}, {Roy}, {Ray}, {Gupta},
  {Bhattacharya}, {Romani}, {Ransom}, {Ferrara}, {Wolff}, {Camilo}, {Cognard},
  {Harding}, {den Hartog}, {Johnston}, {Keith}, {Kerr}, {Michelson}, {Saz
  Parkinson}, {Wood}, \& {Wood}}]{2013ApJ...773L..12B}
{Bhattacharyya}, B., {Roy}, J., {Ray}, P.~S., {et~al.} 2013, \apjl, 773, L12,
  \dodoi{10.1088/2041-8205/773/1/L12}

\bibitem[{{Bock} {et~al.}(1999){Bock}, {Large}, \&
  {Sadler}}]{1999AJ....117.1578B}
{Bock}, D.~C.-J., {Large}, M.~I., \& {Sadler}, E.~M. 1999, \aj, 117, 1578,
  \dodoi{10.1086/300786}

\bibitem[{{Breton} {et~al.}(2013){Breton}, {van Kerkwijk}, {Roberts},
  {Hessels}, {Camilo}, {McLaughlin}, {Ransom}, {Ray}, \& {Stairs}}]{Breton2012}
{Breton}, R.~P., {van Kerkwijk}, M.~H., {Roberts}, M.~S.~E., {et~al.} 2013,
  \apj, 769, 108, \dodoi{10.1088/0004-637X/769/2/108}

\bibitem[{{Bruel}(2019)}]{2019A&A...622A.108B}
{Bruel}, P. 2019, \aap, 622, A108, \dodoi{10.1051/0004-6361/201834555}

\bibitem[{{Bruel} {et~al.}(2018){Bruel}, {Burnett}, {Digel}, {Johannesson},
  {Omodei}, \& {Wood}}]{Bruel2018+P305}
{Bruel}, P., {Burnett}, T.~H., {Digel}, S.~W., {et~al.} 2018, arXiv e-prints,
  arXiv:1810.11394.
\newblock \doarXiv{1810.11394}

\bibitem[{{Buccheri} {et~al.}(1987){Buccheri}, {Sacco}, \&
  {Ozel}}]{1987A&A...175..353B}
{Buccheri}, R., {Sacco}, B., \& {Ozel}, M.~E. 1987, \aap, 175, 353

\bibitem[{{Camilo} {et~al.}(2015){Camilo}, {Kerr}, {Ray}, {Ransom},
  {Sarkissian}, {Cromartie}, {Johnston}, {Reynolds}, {Wolff}, {Freire},
  {Bhattacharyya}, {Ferrara}, {Keith}, {Michelson}, {Saz Parkinson}, \&
  {Wood}}]{2015ApJ...810...85C}
{Camilo}, F., {Kerr}, M., {Ray}, P.~S., {et~al.} 2015, \apj, 810, 85,
  \dodoi{10.1088/0004-637X/810/2/85}

\bibitem[{{Clark} {et~al.}(2017){Clark}, {Wu}, {Pletsch}, {Guillemot}, {Allen},
  {Aulbert}, {Beer}, {Bock}, {Cu{\'e}llar}, {Eggenstein}, {Fehrmann}, {Kramer},
  {Machenschalk}, \& {Nieder}}]{clark2017}
{Clark}, C.~J., {Wu}, J., {Pletsch}, H.~J., {et~al.} 2017, \apj, 834, 106,
  \dodoi{10.3847/1538-4357/834/2/106}

\bibitem[{{Cognard} {et~al.}(2011){Cognard}, {Guillemot}, {Johnson}, {Smith},
  {Venter}, {Harding}, {Wolff}, {Cheung}, {Donato}, {Abdo}, {Ballet}, {Camilo},
  {Desvignes}, {Dumora}, {Ferrara}, {Freire}, {Grove}, {Johnston}, {Keith},
  {Kramer}, {Lyne}, {Michelson}, {Parent}, {Ransom}, {Ray}, {Romani}, {Saz
  Parkinson}, {Stappers}, {Theureau}, {Thompson}, {Weltevrede}, \&
  {Wood}}]{2011ApJ...732...47C}
{Cognard}, I., {Guillemot}, L., {Johnson}, T.~J., {et~al.} 2011, \apj, 732, 47,
  \dodoi{10.1088/0004-637X/732/1/47}

\bibitem[{{Condon} {et~al.}(1998){Condon}, {Cotton}, {Greisen}, {Yin},
  {Perley}, {Taylor}, \& {Broderick}}]{1998AJ....115.1693C}
{Condon}, J.~J., {Cotton}, W.~D., {Greisen}, E.~W., {et~al.} 1998, \aj, 115,
  1693, \dodoi{10.1086/300337}

\bibitem[{{Cordes} \& {Lazio}(2002)}]{NE2001}
{Cordes}, J.~M., \& {Lazio}, T.~J.~W. 2002.
\newblock \doarXiv{astro-ph/0207156}

\bibitem[{{Cromartie} {et~al.}(2016){Cromartie}, {Camilo}, {Kerr}, {Deneva},
  {Ransom}, {Ray}, {Ferrara}, {Michelson}, \& {Wood}}]{2016ApJ...819...34C}
{Cromartie}, H.~T., {Camilo}, F., {Kerr}, M., {et~al.} 2016, \apj, 819, 34,
  \dodoi{10.3847/0004-637X/819/1/34}

\bibitem[{{de Jager} {et~al.}(1989){de Jager}, {Raubenheimer}, \&
  {Swanepoel}}]{dejager1989}
{de Jager}, O.~C., {Raubenheimer}, B.~C., \& {Swanepoel}, J.~W.~H. 1989, \aap,
  221, 180

\bibitem[{{Draghis} {et~al.}(2019){Draghis}, {Romani}, {Filippenko}, {Brink},
  {Zheng}, {Halpern}, \& {Camilo}}]{Draghis2019+BWs}
{Draghis}, P., {Romani}, R.~W., {Filippenko}, A.~V., {et~al.} 2019, \apj, 883,
  108, \dodoi{10.3847/1538-4357/ab378b}

\bibitem[{{Eggleton}(1983)}]{egg83}
{Eggleton}, P.~P. 1983, \apj, 268, 368, \dodoi{10.1086/160960}

\bibitem[{{Espinoza} {et~al.}(2013){Espinoza}, {Guillemot}, {{\c{C}}elik},
  {Weltevrede}, {Stappers}, {Smith}, {Kerr}, {Zavlin}, {Cognard}, {Eatough},
  {Freire}, {Janssen}, {Camilo}, {Desvignes}, {Hewitt}, {Hou}, {Johnston},
  {Keith}, {Kramer}, {Lyne}, {Manchester}, {Ransom}, {Ray}, {Shannon},
  {Theureau}, \& {Webb}}]{Espinoza2013}
{Espinoza}, C.~M., {Guillemot}, L., {{\c{C}}elik}, {\"O}., {et~al.} 2013,
  \mnras, 430, 571, \dodoi{10.1093/mnras/sts657}

\bibitem[{{Frail} {et~al.}(2016){Frail}, {Mooley}, {Jagannathan}, \&
  {Intema}}]{2016MNRAS.461.1062F}
{Frail}, D.~A., {Mooley}, K.~P., {Jagannathan}, P., \& {Intema}, H.~T. 2016,
  \mnras, 461, 1062, \dodoi{10.1093/mnras/stw1390}

\bibitem[{{Frail} {et~al.}(2018){Frail}, {Ray}, {Mooley}, {Hancock}, {Burnett},
  {Jagannathan}, {Ferrara}, {Intema}, {de Gasperin}, {Demorest}, {Stovall}, \&
  {McKinnon}}]{2018MNRAS.475..942F}
{Frail}, D.~A., {Ray}, P.~S., {Mooley}, K.~P., {et~al.} 2018, \mnras, 475, 942,
  \dodoi{10.1093/mnras/stx3281}

\bibitem[{{Fukugita} {et~al.}(1995){Fukugita}, {Shimasaku}, \&
  {Ichikawa}}]{fuk95}
{Fukugita}, M., {Shimasaku}, K., \& {Ichikawa}, T. 1995, \pasp, 107, 945,
  \dodoi{10.1086/133643}

\bibitem[{{Gaia Collaboration} {et~al.}(2021){Gaia Collaboration}, {Brown},
  {Vallenari}, {Prusti}, {de Bruijne}, {Babusiaux}, {Biermann}, {Creevey},
  {Evans}, {Eyer}, \& et~al.}]{bro21}
{Gaia Collaboration}, {Brown}, A.~G.~A., {Vallenari}, A., {et~al.} 2021, \aap,
  649, A1, \dodoi{10.1051/0004-6361/202039657}

\bibitem[{{He} {et~al.}(2013){He}, {Ng}, \& {Kaspi}}]{HeNgKaspi2013}
{He}, C., {Ng}, C.~Y., \& {Kaspi}, V.~M. 2013, \apj, 768, 64,
  \dodoi{10.1088/0004-637X/768/1/64}

\bibitem[{{Intema} {et~al.}(2017){Intema}, {Jagannathan}, {Mooley}, \&
  {Frail}}]{ijmf17}
{Intema}, H.~T., {Jagannathan}, P., {Mooley}, K.~P., \& {Frail}, D.~A. 2017,
  \aap, 598, A78, \dodoi{10.1051/0004-6361/201628536}

\bibitem[{{Johnson} {et~al.}(2014){Johnson}, {Venter}, {Harding}, {Guillemot},
  {Smith}, {Kramer}, {{\c{C}}elik}, {den Hartog}, {Ferrara}, {Hou}, {Lande}, \&
  {Ray}}]{Johnson2014}
{Johnson}, T.~J., {Venter}, C., {Harding}, A.~K., {et~al.} 2014, \apjs, 213, 6,
  \dodoi{10.1088/0067-0049/213/1/6}

\bibitem[{{Kaplan} {et~al.}(2016){Kaplan}, {Kupfer}, {Nice}, {Irrgang},
  {Heber}, {Arzoumanian}, {Beklen}, {Crowter}, {DeCesar}, {Demorest}, {Dolch},
  {Ellis}, {Ferdman}, {Ferrara}, {Fonseca}, {Gentile}, {Jones}, {Jones},
  {Kreuzer}, {Lam}, {Levin}, {Lorimer}, {Lynch}, {McLaughlin}, {Miller}, {Ng},
  {Pennucci}, {Prince}, {Ransom}, {Ray}, {Spiewak}, {Stairs}, {Stovall},
  {Swiggum}, \& {Zhu}}]{Kaplan2016}
{Kaplan}, D.~L., {Kupfer}, T., {Nice}, D.~J., {et~al.} 2016, \apj, 826, 86,
  \dodoi{10.3847/0004-637X/826/1/86}

\bibitem[{{Kerr}(2011)}]{kerr2011}
{Kerr}, M. 2011, \apj, 732, 38, \dodoi{10.1088/0004-637X/732/1/38}

\bibitem[{{Kerr} {et~al.}(2015){Kerr}, {Ray}, {Johnston}, {Shannon}, \&
  {Camilo}}]{kerr2015b}
{Kerr}, M., {Ray}, P.~S., {Johnston}, S., {Shannon}, R.~M., \& {Camilo}, F.
  2015, \apj, 814, 128, \dodoi{10.1088/0004-637X/814/2/128}

\bibitem[{{Kerr} {et~al.}(2012){Kerr}, {Camilo}, {Johnson}, {Ferrara},
  {Guillemot}, {Harding}, {Hessels}, {Johnston}, {Keith}, {Kramer}, {Ransom},
  {Ray}, {Reynolds}, {Sarkissian}, \& {Wood}}]{2012ApJ...748L...2K}
{Kerr}, M., {Camilo}, F., {Johnson}, T.~J., {et~al.} 2012, \apjl, 748, L2,
  \dodoi{10.1088/2041-8205/748/1/L2}

\bibitem[{{Luo} {et~al.}(2021){Luo}, {Ransom}, {Demorest}, {Ray}, {Archibald},
  {Kerr}, {Jennings}, {Bachetti}, {van Haasteren}, {Champagne}, {Colen},
  {Phillips}, {Zimmerman}, {Stovall}, {Lam}, \& {Jenet}}]{PINT}
{Luo}, J., {Ransom}, S., {Demorest}, P., {et~al.} 2021, \apj, 911, 45,
  \dodoi{10.3847/1538-4357/abe62f}

\bibitem[{{Manchester} {et~al.}(2005){Manchester}, {Hobbs}, {Teoh}, \&
  {Hobbs}}]{manchester2005}
{Manchester}, R.~N., {Hobbs}, G.~B., {Teoh}, A., \& {Hobbs}, M. 2005, \aj, 129,
  1993, \dodoi{10.1086/428488}

\bibitem[{{Nieder} {et~al.}(2020){Nieder}, {Allen}, {Clark}, \&
  {Pletsch}}]{Nieder2020+Methods}
{Nieder}, L., {Allen}, B., {Clark}, C.~J., \& {Pletsch}, H.~J. 2020, \apj, 901,
  156, \dodoi{10.3847/1538-4357/abaf53}

\bibitem[{{Nieder} {et~al.}(2019){Nieder}, {Clark}, {Bassa}, {Wu}, {Singh},
  {Donner}, {Allen}, {Breton}, {Dhillon}, {Eggenstein}, {Hessels}, {Kennedy},
  {Kerr}, {Littlefair}, {Marsh}, {Mata S{\'a}nchez}, {Papa}, {Ray}, {Steltner},
  \& {Verbiest}}]{nieder2019}
{Nieder}, L., {Clark}, C.~J., {Bassa}, C.~G., {et~al.} 2019, \apj, 883, 42,
  \dodoi{10.3847/1538-4357/ab357e}

\bibitem[{{Owen}(1996)}]{owen1996}
{Owen}, B.~J. 1996, \prd, 53, 6749, \dodoi{10.1103/PhysRevD.53.6749}

\bibitem[{{Pletsch} \& {Clark}(2014)}]{pletsch2014}
{Pletsch}, H.~J., \& {Clark}, C.~J. 2014, \apj, 795, 75,
  \dodoi{10.1088/0004-637X/795/1/75}

\bibitem[{{Ransom} {et~al.}(2002){Ransom}, {Eikenberry}, \&
  {Middleditch}}]{2002AJ....124.1788R}
{Ransom}, S.~M., {Eikenberry}, S.~S., \& {Middleditch}, J. 2002, \aj, 124,
  1788, \dodoi{10.1086/342285}

\bibitem[{{Ransom} {et~al.}(2011){Ransom}, {Ray}, {Camilo}, {Roberts},
  {{\c{C}}elik}, {Wolff}, {Cheung}, {Kerr}, {Pennucci}, {DeCesar}, {Cognard},
  {Lyne}, {Stappers}, {Freire}, {Grove}, {Abdo}, {Desvignes}, {Donato},
  {Ferrara}, {Gehrels}, {Guillemot}, {Gwon}, {Harding}, {Johnston}, {Keith},
  {Kramer}, {Michelson}, {Parent}, {Saz Parkinson}, {Romani}, {Smith},
  {Theureau}, {Thompson}, {Weltevrede}, {Wood}, \&
  {Ziegler}}]{2011ApJ...727L..16R}
{Ransom}, S.~M., {Ray}, P.~S., {Camilo}, F., {et~al.} 2011, \apjl, 727, L16,
  \dodoi{10.1088/2041-8205/727/1/L16}

\bibitem[{{Ray} {et~al.}(2011){Ray}, {Kerr}, {Parent}, {Abdo}, {Guillemot},
  {Ransom}, {Rea}, {Wolff}, {Makeev}, {Roberts}, {Camilo}, {Dormody}, {Freire},
  {Grove}, {Gwon}, {Harding}, {Johnston}, {Keith}, {Kramer}, {Michelson},
  {Romani}, {Saz Parkinson}, {Thompson}, {Weltevrede}, {Wood}, \&
  {Ziegler}}]{ray2011}
{Ray}, P.~S., {Kerr}, M., {Parent}, D., {et~al.} 2011, \apjs, 194, 17,
  \dodoi{10.1088/0067-0049/194/2/17}

\bibitem[{{Ray} {et~al.}(2012){Ray}, {Abdo}, {Parent}, {Bhattacharya},
  {Bhattacharyya}, {Camilo}, {Cognard}, {Theureau}, {Ferrara}, {Harding},
  {Thompson}, {Freire}, {Guillemot}, {Gupta}, {Roy}, {Hessels}, {Johnston},
  {Keith}, {Shannon}, {Kerr}, {Michelson}, {Romani}, {Kramer}, {McLaughlin},
  {Ransom}, {Roberts}, {Saz Parkinson}, {Ziegler}, {Smith}, {Stappers},
  {Weltevrede}, \& {Wood}}]{PSC}
{Ray}, P.~S., {Abdo}, A.~A., {Parent}, D., {et~al.} 2012, in 2011 Fermi
  Symposium proceedings - eConf C110509, arXiv:1205.3089.
\newblock \doarXiv{1205.3089}

\bibitem[{{Ray} {et~al.}(2013){Ray}, {Ransom}, {Cheung}, {Giroletti},
  {Cognard}, {Camilo}, {Bhattacharyya}, {Roy}, {Romani}, {Ferrara},
  {Guillemot}, {Johnston}, {Keith}, {Kerr}, {Kramer}, {Pletsch}, {Saz
  Parkinson}, \& {Wood}}]{2013ApJ...763L..13R}
{Ray}, P.~S., {Ransom}, S.~M., {Cheung}, C.~C., {et~al.} 2013, \apjl, 763, L13,
  \dodoi{10.1088/2041-8205/763/1/L13}

\bibitem[{{Ray} {et~al.}(2019){Ray}, {Guillot}, {Ransom}, {Kerr}, {Bogdanov},
  {Harding}, {Wolff}, {Malacaria}, {Gendreau}, {Arzoumanian}, {Markwardt},
  {Soong}, \& {Doty}}]{rgr+19}
{Ray}, P.~S., {Guillot}, S., {Ransom}, S.~M., {et~al.} 2019, \apjl, 878, L22,
  \dodoi{10.3847/2041-8213/ab2539}

\bibitem[{{Schlafly} \& {Finkbeiner}(2011)}]{sch11}
{Schlafly}, E.~F., \& {Finkbeiner}, D.~P. 2011, \apj, 737, 103,
  \dodoi{10.1088/0004-637X/737/2/103}

\bibitem[{{Schwarz}(1978)}]{schwarz1978}
{Schwarz}, G. 1978, Annals of Statistics, 6, 461

\bibitem[{{Stovall} {et~al.}(2014){Stovall}, {Lynch}, {Ransom}, {Archibald},
  {Banaszak}, {Biwer}, {Boyles}, {Dartez}, {Day}, {Ford}, {Flanigan}, {Garcia},
  {Hessels}, {Hinojosa}, {Jenet}, {Kaplan}, {Karako-Argaman}, {Kaspi},
  {Kondratiev}, {Leake}, {Lorimer}, {Lunsford}, {Martinez}, {Mata},
  {McLaughlin}, {Roberts}, {Rohr}, {Siemens}, {Stairs}, {van Leeuwen},
  {Walker}, \& {Wells}}]{2014ApJ...791...67S}
{Stovall}, K., {Lynch}, R.~S., {Ransom}, S.~M., {et~al.} 2014, \apj, 791, 67,
  \dodoi{10.1088/0004-637X/791/1/67}

\bibitem[{{Strader} {et~al.}(2019){Strader}, {Swihart}, {Chomiuk}, {Bahramian},
  {Britt}, {Cheung}, {Dage}, {Halpern}, {Li}, {Mignani}, {Orosz}, {Peacock},
  {Salinas}, {Shishkovsky}, \& {Tremou}}]{Strader2019}
{Strader}, J., {Swihart}, S., {Chomiuk}, L., {et~al.} 2019, \apj, 872, 42,
  \dodoi{10.3847/1538-4357/aafbaa}

\bibitem[{{Tauris} \& {Savonije}(1999)}]{TS99}
{Tauris}, T.~M., \& {Savonije}, G.~J. 1999, \aap, 350, 928.
\newblock \doarXiv{astro-ph/9909147}

\bibitem[{{van Straten} \& {Bailes}(2011)}]{dspsr}
{van Straten}, W., \& {Bailes}, M. 2011, \pasa, 28, 1, \dodoi{10.1071/AS10021}

\bibitem[{{Yao} {et~al.}(2017){Yao}, {Manchester}, \& {Wang}}]{YMW16}
{Yao}, J.~M., {Manchester}, R.~N., \& {Wang}, N. 2017, \apj, 835, 29,
  \dodoi{10.3847/1538-4357/835/1/29}

\end{thebibliography}

\end{document}